# Prevalence of SED Turndown among Classical Be Stars: Are All Be Stars Close Binaries?

Robert Klement[1], A. C. Carciofi[2], T. Rivinius[3], R. Ignace[4], L. D. Matthews[5], K. Torstensson[3], D. Gies[6], R. G. Vieira[2], N. D. Richardson[7], A. Domiciano de Souza[8], J. E. Bjorkman[7], G. Hallinan[9], D. M. Faes[2], B. Mota[2], A. D. Gullingsrud[7], C. de Breuck[10], P. Kervella[11], M. Curé[12], and D. Gunawan[12]

[1] The CHARA Array of Georgia State University, Mount Wilson Observatory, Mount Wilson, CA 91023, USA; robertklement@gmail.com
[2] Instituto de Astronomia, Geofísica e Ciências Atmosféricas, Universidade de São Paulo, Rua do Matão 1226, Cidade Universitária, 05508-900 São Paulo, SP, Brazil
[3] European Southern Observatory, Alonso de Córdova 3107, Vitacura, Casilla 19001, Santiago de Chile, Chile
[4] Department of Physics & Astronomy, East Tennessee State University, Johnson City, TN 37614, USA
[5] Massachusetts Institute of Technology Haystack Observatory, 99 Millstone Rd., Westford, MA 01886, USA
[6] Center for High Angular Resolution Astronomy and Department of Physics and Astronomy, Georgia State University, P.O. Box 5060, Atlanta, GA 30302-5060, USA
[7] Ritter Observatory, Department of Physics & Astronomy, University of Toledo, Toledo, OH 43606, USA
[8] Université Côte d'Azur, Observatoire de la Côte d'Azur, CNRS, UMR7293 Lagrange, 28 Av. Valrose, F-06108 Nice Cedex 2, France
[9] Cahill Center for Astrophysics, California Institute of Technology, 1200 E. California Blvd., MC 249-17, Pasadena, CA 91125, USA
[10] European Southern Observatory, Karl-Schwarzschild-Str. 2, D-85748 Garching bei München, Germany
[11] LESIA, Observatoire de Paris, Université PSL, CNRS, Sorbonne Université, Univ. Paris Diderot, Sorbonne Paris Cité, 5 place JulesJanssen, F-92195 Meudon, France
[12] Instituto de Física y Astronomía, Facultad de Ciencias, Universidad de Valparaíso, Casilla 5030, Valparaíso, Chile

## Abstract

Rapid rotation is a fundamental characteristic of classical Be stars and a crucial property allowing for the formation of their circumstellar disks. Past evolution in a mass and angular momentum transferring binary system offers a plausible solution to how Be stars attained their fast rotation. Although the subdwarf remnants of mass donors in such systems should exist in abundance, only a few have been confirmed due to tight observational constraints. An indirect method of detecting otherwise hidden companions is offered by their effect on the outer parts of Be star disks, which are expected to be disrupted or truncated. In the context of the infrared and radio continuum excess radiation originating in the disk, the disk truncation can be revealed by a turndown in the spectral energy distribution due to reduced radio flux levels. In this work, we search for signs of spectral turndown in a sample of 57 classical Be stars with radio data, which include new data for 23 stars and the longest-wavelength detections so far ($\lambda \approx 10$ cm) for two stars. We confidently detect the turndown for all 26 stars with sufficient data coverage (20 of which are not known to have close binary companions). For the remaining 31 stars, the data are inconclusive as to whether the turndown is present or not. The analysis suggests that many if not all Be stars have close companions influencing their outer disks. If confirmed to be subdwarf companions, the mass transfer spin-up scenario might explain the existence of the vast majority of classical Be stars.

*Unified Astronomy Thesaurus concepts:* Be stars (142); Circumstellar disks (235); Binary stars (154); Radio continuum emission (1340); Early-type emission stars (428); Stellar rotation (1629); Stellar evolution (1599); Infrared excess (788); Stellar astronomy (1583); Circumstellar matter (241); Radio astronomy (1338); Subdwarf stars (2054)

*Supporting material:* figure sets

## 1. Introduction

Classical Be stars (CBes) are rapidly rotating B stars showing line emission, which originates from their self-ejected circumstellar disks (Struve 1931; Rivinius et al. 2013). Although the mass ejection mechanism is not fully constrained thus far, the physics of the disk material itself, once in orbit, is governed by viscosity (Shakura & Sunyaev 1973; Martin et al. 2019). Observational evidence points at rather high values of the viscosity parameter $\alpha$ of the order of a few tenths (Carciofi et al. 2012; Ghoreyshi et al. 2018; Rímulo et al. 2018). The physical model that so far best describes these disks is the so-called viscous decretion disk (VDD) model (Lee et al. 1991; Carciofi 2011).

Under the condition of a sufficient and steady mass ejection rate, the VDD expands outward and is expected to eventually settle into a steady state with a characteristic density, velocity, and ionization distribution: the radial decrease in density is close to a power law, the disk rotates close to Keplerian while only slowly diffusing outward, and the gas in the disk is at least partially ionized. Geometrically, the disk is thin close to the star but flares at larger distances. Assuming a purely hydrogen composition, the temperature structure of the disk is expected to be mostly isothermal, with the exception of a temperature dip appearing in the vicinity of the star (Carciofi & Bjorkman 2006, 2008). However, models with a more realistic chemical mixing process suggest that the temperature structure may be more complex (Jones et al. 2004; McGill et al. 2013).

The characteristic observables of CBes are (1) line emission, which is the strongest in recombination lines of hydrogen such as H$\alpha$; (2) continuum excess due to free–free and bound–free disk emission (mostly in the infrared (IR) and radio); and (3) linear continuum polarization due to scattering off free electrons in the inner parts of the disk. The CBes and their disks are known to be variable on many timescales ranging from a few minutes to decades. The sources of the observable variations include nonradial pulsations of the central star, density waves propagating through the disk, varying mass-loss rates, and binarity (Rivinius et al. 2013). Perhaps the most striking feature of many CBes is the fact that the line emission and circumstellar continuum, and therefore the circumstellar disk, can disappear entirely over the course of months to decades (e.g., π Aqr; Bjorkman et al. 2002) or start growing from scratch (e.g., ω Ori; Sonneborn et al. 1988). With the advent of large-scale robotic photometric surveys, these episodes of disk formation and dissipation are now documented by the thousands (e.g., Keller et al. 2002; Sabogal et al. 2005; Labadie-Bartz et al. 2018; Rímulo et al. 2018). Some stars show almost periodic or quasi-periodic phases of disk growth and dissipation (e.g., ω CMa; Ghoreyshi et al. 2018), but in most of them, this seems to happen randomly. The CBes known for such variable disks are often referred to as *active* Be stars, while *stable* CBes are observed to have disks with very little small-scale and no large-scale variation.

Rapid rotation is an essential property allowing for the formation of the disk, although it is probably not the only one. The other contributing mechanism is likely a combination of nonradial pulsations and small-scale magnetic fields (Rivinius et al. 2013). The origin of the rapid rotation in CBes is still a matter of active debate. Often-suggested scenarios include CBes being born as fast rotators (Zorec & Briot 1997), spinning up during main-sequence (MS) evolution due to contraction of the hydrogen-burning core (Ekström et al. 2008; Granada et al. 2013), and past evolution in a mass-transferring binary system (Rappaport & van den Heuvel 1982; Pols et al. 1991; Gies 2000; Shao & Li 2014) or a binary merger (Vanbeveren & Mennekens 2017). In the mass-transferring binary scenario, the initially more massive star is the first to evolve beyond the MS, becoming the mass donor by Roche lobe overflow. The mass gainer accretes the overflowing material, spins up due to angular momentum transfer, and becomes the more massive CBe component. The remnant of the mass donor then becomes a black hole (BH), neutron star (NS), white dwarf (WD), or subdwarf O or B star (sdOB), depending on the progenitor mass. Observational evidence seems to favor this scenario as the most likely origin of the rapid rotation of CBes (McSwain & Gies 2005; Schootemeijer et al. 2018).

The present work is an observational investigation of the possible prevalence of (unseen) companions that could have played a role in the past spin-up of CBes. We analyze the spectral energy distributions (SEDs) of 57 CBes for which radio flux density measurements (detections or upper limits) are available in order to search for effects of possible outer disk truncation caused by the orbiting companion. In doing so, we also present a full set of radio measurements of CBes that includes both archival and new, previously unpublished measurements.

The paper is structured as follows. In Section 2, we provide more theoretical background; in Section 3 we describe in detail the sample of the studied stars; in Section 4 we describe the observational data set; and in Section 5 we discuss the properties of the observed SEDs, focusing on the incidence of SED turndown. We discuss the findings in Section 6 before concluding in Section 7.

## 2. Binarity and SED Turndown

### 2.1. Binarity among CBes

Most massive stars are known to be components of multiple stellar systems (see, e.g., Sana 2017, for a recent overview). Among B-type stars in the Large Magellanic Cloud, the fraction of spectroscopic binaries in 408 studied systems was reported to be $0.58 \pm 0.11$ (Dunstall et al. 2015). Consequently, CBes can also be expected to reside in binary systems.

It was suggested early on that CBes may be binaries with ongoing mass transfer, where the characteristic line emission originates in an accretion disk (Kříž & Harmanec 1975). While systems like this do exist, such as the archetype example β Lyr (Mourard et al. 2018) and Algol-type variables (e.g., CX Dra; Richards et al. 2000), CBes as defined above constitute a physically different class of stars (see Section 1.1.1 of Rivinius et al. 2013). No CBe has been found to have a Roche lobe filling companion.

For the purposes of this study, we are interested in close binary systems containing a CBe and a companion that tidally influences the outer parts of the CBe disk. As the companions in this case are likely to accrete at least some of the disk material, there is an ongoing mass transfer between the components. This present-day mass transfer is not connected to the possible past mass transfer that caused the spin-up of the CBe, and close companions defined in this way can therefore include both remnants of past mass transfer and companions not necessarily involved in the past CBe spin-up. Such close binary systems have orbital periods of a few tens to several hundred days corresponding to binary separation of a few tens to a few hundred radii of the primary B star. At distances of >100 pc, typical of most bright CBes, the on-sky angular separations are 10 mas or less.

Binary systems that experienced past mass transfer and a spin-up of the present-day CBe component include Be X-ray type binaries (BeXRBs) and Be+sdOB systems. The BeXRBs consist of CBes and compact, post-supernova objects (BHs, NSs, or WDs) and are sources of characteristic X-ray radiation, which makes them fairly easy to detect (Reig 2011). Mainly CBe+NS systems were detected so far (81 according to Shao & Li 2014), while WD companions, although thought to be common, still remain elusive. Thought to be rare and hard to detect using conventional X-ray surveys, only one CBe+BH system was confirmed so far (Casares et al. 2014). Most BeXRBs have eccentric orbits caused by the kick from the past supernova explosion of the present compact component. It is thought that Be+sdOB systems, in which the sdOB components are the remaining cores of stars that completely lost their outer shells via Roche lobe overflow onto the present CBe, exist in abundance (Rivinius et al. 2013; Shao & Li 2014). Although only five have been directly detected so far, the number of potential candidates is much larger (Wang et al. 2018). The sdOB stars have typical masses of $1\,M_\odot$ and temperatures up to ~50,000 K.

Probably the most widespread way to detect close binaries in general is through periodic radial velocity (RV) variations of the binary components as they orbit each other, which can be detected in the composite spectra. This method requires the orbital plane to be inclined with respect to the observer, and it also requires the companion to be massive enough to cause detectable RV shifts in the given orbital configuration. The latter requirement is especially important for the case of CBes, in which spectral lines are rotationally broadened, making it difficult to detect low-amplitude RV shifts ($\lesssim 10$ km s$^{-1}$) caused by low-mass companions. Another method is provided by optical long-baseline interferometry (OLBI), which is capable of direct detection of companions down to submilliarcsecond angular resolution, provided that the companion is bright enough compared to the primary CBe (Mourard et al. 2015).

An indirect method of detecting close companions that can otherwise remain elusive is the radio SED turndown caused by disk "truncation" through the gravitational influence of an orbiting companion (Klement et al. 2015, 2017b; also see the next section). Throughout this work, we will use the term truncation to refer to any kind of disruption of the outer disk that has observable effects on the radio SED. The advantage of this method as compared to the other methods is that it does not require the orbit to be inclined with respect to the observer, and the companion can be arbitrarily faint and of very low mass. This method also has the potential of detecting wider binaries than RV methods can.

The nature of sdOB companions provides us with other binary detection methods specific to the class of Be+sdOB binaries. Under favorable circumstances, the photospheric spectral signatures of hot sdOB companions can be directly detected in the far-UV domain, observable from space by, e.g., the *International Ultraviolet Explorer* (*IUE*) and the *Hubble Space Telescope* (*HST*). To date, five CBe+sdOB binaries have been confirmed using this method, namely, $\varphi$ Per (Thaller et al. 1995), FY CMa (Peters et al. 2008), 59 Cyg (Peters et al. 2013), HR 2142 (Peters et al. 2016), and 60 Cyg (Wang et al. 2017). Using the same method, eight more were classified as candidate Be+sdOB binaries and four more as potential candidates (Wang et al. 2018).

Since sdOB stars are hotter than CBes, their radiation can heat up the outer parts of the CBe disk facing the companion, which can lead to orbital phase-locked spectral features, such as emission in specific spectral lines requiring hotter conditions. The CBes for which such behavior was detected include, e.g., $o$ Pup (Koubský et al. 2012a), HD 161306 (Koubský et al. 2014), and HD 55606 (Chojnowski et al. 2018), which were all proposed to be Be+sdOB systems. The companions also likely accrete part of the outer disk material, thus forming a secondary accretion disk that can be the origin of high-excitation wind-forming lines (as observed in, e.g., $\beta$ CMi; Klement et al. 2015).

Several other spectral signatures have been linked to possible binarity in CBes. These include triple-peaked and otherwise peculiar profiles of the H$\alpha$ line (Section 6.1 of Rivinius et al. 2013), possibly caused by the disk being perturbed by the companion, and strong metallic emission such as that of the IR Ca II triplet (Polidan 1976; Jeffery & Pollacco 1998; Koubský et al. 2011) coinciding with very little or no Paschen line emission (Koubský et al. 2012b; Klement et al. 2015).

### 2.2. SED of CBes in IR and Radio

The circumstellar envelopes of CBes are highly ionized and devoid of a dust component. The ionized gas reprocesses part of the stellar radiation via bound–free and, more importantly, free–free emission mechanisms, thus causing continuum excess emission over the flux originating from the stellar photosphere mostly at radio and IR wavelengths (Panagia & Felli 1975; Wright & Barlow 1975; Lamers & Waters 1984). The amount of excess continuum emission rises with increasing wavelength due to the fact that hydrogen free–free opacity is proportional to $\lambda^2$ (e.g., Lamers & Waters 1984; Vieira et al. 2015). For envelopes of sufficient sizes and densities, the bound–free and free–free components can consequently become a major contribution to the SED already in the visible and near-IR, and from the mid-IR longward, the free–free emission begins to entirely dominate over the photospheric flux. The IR and radio SED is therefore a crucial probe of the density and velocity structure of the gaseous envelopes surrounding early-type stars such as CBes.

The observed IR SEDs of hot stars with gaseous envelopes show spectral slopes consistent with a simple power law $F_\nu \propto \lambda^{-\kappa}$, with $\kappa$ having intermediate values between $-0.1$ (optically thin homogeneous plasma such as low-density gas) and 2.0 (optically thick plasma such as stellar photospheres). The intermediate values are due to the inner parts of the gaseous envelope becoming partially optically thick with increasing wavelength, as well as having an inhomogeneous density distribution (e.g., Wright & Barlow 1975). While spherically symmetric isothermal envelopes with density distribution $\propto r^{-2}$ (stellar winds) are expected to have $\kappa \sim 0.6$, circumstellar disks of CBes are observed to typically have $\kappa \sim 0.8$–1.2 in the IR, which is compatible with the radial density exponent $n$ in $\rho \propto r^{-n}$ being between 2.5 and 3.25 (Cote & Waters 1987). Throughout this work, we adopt the $F_\nu \propto \lambda^{-\kappa}$ convention used in past studies of the IR and radio excesses of CBes.

Vieira et al. (2015) derived an approximate formula for the dependence of the spectral slope $\alpha_{IR}$ on the density slope exponent $n$ for the case of CBes, where $\alpha_{IR}$ is the spectral slope derived from $F_\lambda \propto \lambda^{-\alpha_{IR}}$. The spectral slopes $\kappa$ and $\alpha_{IR}$ are then simply related as $\kappa = \alpha_{IR} - 2$; therefore,

$$\alpha_{IR} = \kappa + 2 = 4 - \frac{4+2u}{2n-\beta}, \quad (1)$$

where $u$ is defined as

$$u = \frac{d\ln(g+b)}{d\ln\lambda}, \quad (2)$$

and $\beta$ is the disk flaring exponent. The quantities $g$ and $b$ are temperature- and wavelength-dependent expressions for the free–free and bound–free Gaunt factors, respectively (see Appendix A of Vieira et al. 2015):

$$g(\lambda, T) \simeq \exp[G_0(T) + G_1(T)\ln\lambda + G_2(T)(\ln\lambda)^2], \quad (3)$$

$$b(\lambda, T) \simeq \exp[B_0(T) + B_1(T)\ln\lambda + B_2(T)(\ln\lambda)^2]. \quad (4)$$

The values of the $G$ and $B$ parameters for six representative temperatures are listed in Table A1 of Vieira et al. (2015).

The most important parameter describing the continuum emission of CBes is the size of the optically thick region along the line of sight—the so-called pseudo-photosphere—at the given wavelength (Vieira et al. 2015). In the general case, a typical CBe disk is comprised of both a pseudo-photosphere and an optically thin outer region, with the pseudo-photosphere dominating the observed flux. The two limiting cases are a tenuous (optically thin) disk and a truncated pseudo-photosphere, with the occurrence of these (at the given wavelength) based upon the interplay between the disk density and physical size. The radius of the pseudo-photosphere depends on the wavelength: while at the visible and near-IR, it can be very small or nonexistent, at the mid-IR, it is typically of the order of a few stellar radii. At radio wavelengths where the stars are still bright enough to be detected ($\lambda \sim 10$ cm), it can reach a size on the order of tens of stellar radii (Klement et al. 2015; Vieira et al. 2015). Considering the possible truncation or any kind of change of density structure of the outer parts of CBe disks (beyond ∼10 stellar radii) by orbiting companions, it is therefore the radio part of the SED that is most affected.

### 2.3. Outer Disks of CBes and Radio Observations

In the case of no outer influence on the disk structure, VDDs are expected to extend to a distance of several hundred stellar radii to a point where the disk material attains supersonic velocities in the radial direction and dissolves into the interstellar material (Krtička et al. 2011; Kurfürst et al. 2014). In typical CBe disks, this effect would be seen at wavelengths ≳10 cm (Okazaki 2001; Klement et al. 2017b), where radio observations of stellar objects are generally not feasible due to confusion with other sources and insufficient sensitivity.

Close binary companions, such as those that could be involved in past mass transfer, would truncate the disks closer to the star. Assuming a VDD in a steady state with a density distribution consistent with a single power-law throughout the disk, this effect is observable at millimeter to centimeter regions of the SED. If the disk is fully truncated at a certain distance, the SED is expected to show a turndown at the wavelength at which the size of the pseudo-photosphere matches the size of the disk (Vieira et al. 2015). In this case, the spectral slope longward of the turndown is expected to become parallel to the photospheric spectrum, i.e., to attain a spectral slope corresponding to $\kappa \sim 2$ (Figure 11 of Vieira et al. 2015).

Hydrodynamic simulations of the effect of a close companion on CBe disks predict a steep turndown in the density profile at a truncation radius corresponding to the 3:1 orbital resonance with the binary orbit (Okazaki et al. 2002; Panoglou et al. 2016) and a possible accumulation effect causing enhanced density and a shallower density slope inward of the truncation radius. The simulations presented in these papers were set up in such a way that the disk particles are destroyed once they enter the Roche lobe radius of the secondary, thus not allowing for the possibility that the disk can extend beyond the binary orbit. One can hypothesize that gas from a dense disk would not be able to fully accrete onto the companion, which could lead to the formation of a disk composed of three parts: (1) circumprimary decretion disk, (2) circumsecondary accretion disk, and (3) circumbinary decretion disk. Throughout this work, we will use the term "circumbinary disk" to refer to a disk composed of all three of these parts. In any case, the slope of the radio SED is indicative of the nature of the truncation, and with sufficient data coverage, one should be able to determine whether the disk is fully truncated or not.

The fact that CBes observed at radio wavelengths typically show a turndown between IR and radio SED was noticed for a few stars in the pioneering works of Taylor et al. (1987, 1990), who presented the first flux density measurements of CBes at centimeter wavelengths. Around the same time, CBes were first detected at millimeter wavelengths by Waters et al. (1989, 1991). A few more radio detections in the centimeter region were reported by Dougherty et al. (1991) and Clark et al. (1998), with Dougherty & Taylor (1992) reporting the first and only marginal angular resolution of a CBe disk—$\psi$ Per—in the radio.

Recently, the problem was revisited in the context of the VDD model, which led to the conclusion that the most plausible scenario causing the SED turndown is indeed the presence of small (mostly unseen) companions truncating the disks of CBes at a distance of the order of tens of stellar radii from the central star (Klement et al. 2015, 2017b). New radio data also revealed that the slope of the radio SED is too shallow to be explained by sharp truncation or steep density falloff. This leaves the possibility that the density turndown caused by the companion is more gradual than expected or that the outer parts of the disks are circumbinary, with the companion unable to fully truncate the disk. Based on the truncation radius derived for $\psi$ Per while assuming a sharp truncation, Klement et al. (2017b) argued that the angular resolution of Dougherty & Taylor (1992) was possible only if the disk stretches beyond the orbit of the supposed binary. Evidence for a CBe disk extending beyond the binary orbit was also independently found for the case of Be+sdOB binary HR 2142 using a different method (Peters et al. 2016).

The suggested companion truncating the disk in the particular case of $\beta$ CMi (Klement et al. 2015) has since been confirmed by an analysis of RV shifts of the emission wings of the H$\alpha$ line (Dulaney et al. 2017). In the case of $\gamma$ Cas, for which the companion had been known from before, the spectral slope of the radio SED was found to be almost parallel to the slope in the mid-IR, although the radio fluxes were found to be clearly below the extrapolation of the mid-IR fluxes, thus giving further support to the circumbinary scenario (Klement et al. 2017b). Overall, the available radio data show that a sharp truncation of CBe disks in close binaries appears not to be the general case.

### 3. The Sample

The target list studied in this paper represents the full sample of 57 CBes observed at millimeter and/or centimeter wavelengths and includes both detections and upper limits. The stars are listed in Table 1 along with V magnitudes, spectral types extracted from the Bright Star Catalog (Hoffleit & Jaschek 1991), distances from both Hipparcos (van Leeuwen 2007) and the Gaia DR2 (Gaia Collaboration et al. 2018), information about the radio data that are available, and the number of the figure set in which the SED plot of the corresponding star is included.

#### 3.1. Known Binary CBes in the Sample

Out of the entire sample listed in Table 1, about half of the stars have an entry in the Washington Double Star Catalog, with the vast majority having separations of the order of arcseconds (Mason et al. 2001). Secondary components separated by ≳0″03 are not expected to affect the disks of the primary CBes (see Sections 2.1 and 2.3) and are therefore treated as single stars in our SED analysis. However, companions separated by ≲10″ might contaminate flux measurements by the IR and single-dish radio telescopes due to the large angular response on the sky at longer wavelengths. This will be kept in mind when compiling the SED data (Section 4).

##### 3.1.1. Close Binaries

The known spectroscopic binary CBes contained in our sample are listed in Table 2 along with their orbital period, eccentricity, and physical nature of the companion, when available. The reference in the table corresponds to the latest study of binarity for the given object, with more details on the individual stars given in Appendix A.1.

Table 1
Properties of the CBe Sample

| HR | Name | V Magnitude | Spectral Type | Hipparcos Distance (pc) | Gaia DR2 Distance (pc) | New Radio Data (2010+) | Archival Radio Data | Figure Set |
|---|---|---|---|---|---|---|---|---|
| 193 | $o$ Cas | 4.50 | B5IIIe | $216^{+19}_{-16}$ | $213^{+21}_{-18}$ | ⋯ | VLA | 5 |
| 264 | $\gamma$ Cas | 2.39 | B0IVe | $168.4^{+3.5}_{-3.3}$ | ⋯ | JVLA | JCMT, MRT, VLA | 3 |
| 472 | $\alpha$ Eri | 0.46 | B3Vpe | $42.8^{+1.1}_{-1.0}$ | ⋯ | LABOCA | ⋯ | 6 |
| 496 | $\varphi$ Per | 4.06 | B2Vep | $220.3^{+10.2}_{-9.3}$ | $151^{+12}_{-10}$ | ⋯ | MRT, VLA | 4 |
| 1087 | $\psi$ Per | 4.23 | B5Ve | $178.9^{+7.3}_{-6.8}$ | $168^{+29}_{-21}$ | JVLA | JCMT, MRT, VLA | 3 |
| 1142 | 17 Tau | 3.70 | B6III | $124.1^{+4.0}_{-3.7}$ | $114.9^{+7.4}_{-6.5}$ | ⋯ | VLA | 1 |
| 1165 | $\eta$ Tau | 2.87 | B7IIIe | $123.6^{+6.8}_{-6.1}$ | $125^{+17}_{-14}$ | JVLA | JCMT, MRT, VLA | 3 |
| 1273 | 48 Per | 4.03 | B3Ve | $146.2^{+3.5}_{-3.3}$ | ⋯ | ⋯ | VLA | 3 |
| 1423 | 228 Eri | 5.41 | B1Vne | $465^{+73}_{-55}$ | $468^{+30}_{-27}$ | ⋯ | VLA | 4 |
| 1622 | 11 Cam | 5.08 | B2.5Ve | $210^{+15}_{-13}$ | $216.5^{+7.9}_{-7.3}$ | ⋯ | VLA | 3 |
| 1660 | 105 Tau | 5.92 | B2Ve | $331^{+56}_{-42}$ | $325.1^{+9.9}_{-9.3}$ | ⋯ | VLA | 3 |
| 1789 | 25 Ori | 4.96 | B1Vpe | $318^{+114}_{-67}$ | $257^{+26}_{-22}$ | ⋯ | MRT, VLA | 2 |
| 1910 | $\zeta$ Tau | 3.03 | B4IIIpe | $136^{+17}_{-14}$ | ⋯ | LABOCA, CARMA, JVLA | MRT, VLA | 3 |
| 1934 | $\omega$ Ori | 4.59 | B3IIIe | $424^{+59}_{-46}$ | $316^{+41}_{-32}$ | LABOCA | VLA | 2 |
| 1956 | $\alpha$ Col | 2.65 | B7IVe | $80.1^{+2.4}_{-2.2}$ | $138^{+23}_{-17}$ | LABOCA | ATCA, VLA | 3 |
| 2142 | ⋯ | 5.21 | B2Ven | $403^{+175}_{-94}$ | $471^{+38}_{-33}$ | ⋯ | VLA | 3 |
| 2170 | ⋯ | 5.84 | B4IVe | $312^{+29}_{-24}$ | $336.6^{+9.7}_{-9.1}$ | ⋯ | VLA | 5 |
| 2198 | 69 Ori | 4.92 | B5Vn | $162.1^{+6.8}_{-6.3}$ | $223^{+15}_{-13}$ | ⋯ | MRT | 1 |
| 2249 | HR 2249 | 5.91 | B2.5Vn | $239^{+21}_{-18}$ | $316.7^{+5.7}_{-5.5}$ | ⋯ | MRT | 2 |
| 2356 | $\beta$ Mon A | 4.60 | B3Ve | $207^{+63}_{-39}$ | $191.7^{+9.8}_{-8.9}$ | LABOCA, JVLA | JCMT, MRT, VLA | 3 |
| 2357 | $\beta$ Mon B | 5.00 | B3Ve | $207^{+63}_{-39}$ | $204.3^{+8.9}_{-8.2}$ | JVLA | ⋯ | 2 |
| 2358 | $\beta$ Mon C | 5.32 | B3Ve | $207^{+63}_{-39}$ | $204.3^{+8.9}_{-8.2}$ | JVLA | ⋯ | 3 |
| 2538 | $\kappa$ CMa | 3.89 | B1.5IVne | $202.0^{+5.0}_{-4.8}$ | $211^{+16}_{-14}$ | LABOCA | VLA | 4 |
| 2545 | ⋯ | 5.75 | B6Vnpe | $199^{+12}_{-11}$ | $193.4^{+3.7}_{-3.6}$ | ⋯ | VLA | 2 |
| 2749 | $\omega$ CMa | 3.82 | B2IV-Ve | $279^{+14}_{-13}$ | $205^{+19}_{-16}$ | LABOCA | ⋯ | 6 |
| 2845 | $\beta$ CMi | 2.89 | B8Ve | $49.58^{+0.50}_{-0.49}$ | $49.1^{+3.1}_{-2.8}$ | LABOCA, CARMA, JVLA | JCMT, MRT, VLA | 3 |
| 2911 | OW Pup | 5.44 | B3Vne | $364^{+35}_{-29}$ | $364^{+21}_{-19}$ | ⋯ | VLA | 4 |
| 3034 | $o$ Pup | 4.49 | B0V:pe: | $435^{+48}_{-40}$ | $306^{+30}_{-25}$ | LABOCA | MRT, VLA | 2 |
| 4140 | p Car | 3.27 | B4Vne | $148.1^{+9.3}_{-8.3}$ | $115.2^{+6.5}_{-5.9}$ | LABOCA | ⋯ | 3 |
| 4621 | $\delta$ Cen | 2.52 | B2IVne | $127.2^{+8.1}_{-7.2}$ | ⋯ | LABOCA | ATCA | 3 |
| 4787 | $\kappa$ Dra | 3.89 | B6IIIpe | $150.4^{+8.1}_{-7.3}$ | $140.1^{+6.8}_{-6.2}$ | ⋯ | MRT, VLA | 3 |
| 4897 | $\lambda$ Cru | 4.60 | B4Vne | $117.6^{+3.0}_{-2.8}$ | $109.5^{+3.2}_{-3.0}$ | ⋯ | VLA | 1 |
| 5193 | $\mu$ Cen | 3.43 | B2IV-Ve | $155.0^{+3.9}_{-3.8}$ | $118.5^{+9.5}_{-8.2}$ | ⋯ | ATCA | 5 |
| 5440 | $\eta$ Cen | 2.31 | B1.5Vne | $93.7^{+1.9}_{-1.8}$ | ⋯ | LABOCA | ⋯ | 5 |
| 5551 | $\theta$ Cir | 5.11 | B4Vnpe | $463^{+72}_{-55}$ | $347^{+48}_{-38}$ | ⋯ | ATCA | 2 |
| 5683 | $\mu$ Lup | 4.27 | B8Ve | $102.9^{+8.1}_{-7.0}$ | ⋯ | ⋯ | ATCA | 2 |
| 5941 | 48 Lib | 4.87 | B5IIIp | $143.5^{+5.1}_{-4.8}$ | $133.0^{+4.6}_{-4.3}$ | LABOCA, JVLA | MRT, VLA | 3 |
| 5953 | $\delta$ Sco | 2.32 | B0.3IV | $151^{+23}_{-18}$ | ⋯ | LABOCA, CARMA | ⋯ | 6 |
| 6118 | $\chi$ Oph | 4.43 | B2IV:pe | $161.0^{+6.2}_{-5.8}$ | $122.3^{+4.7}_{-4.4}$ | LABOCA | ATCA, VLA | 3 |
| 6175 | $\zeta$ Oph | 2.56 | O9.5Vn | $112.2^{+2.6}_{-2.5}$ | $172^{+37}_{-26}$ | ⋯ | MRT | 5 |
| 6304 | ⋯ | 6.11 | B2IVne | $431^{+84}_{-61}$ | $663^{+34}_{-30}$ | ⋯ | ATCA | 5 |
| 6451 | $\iota$ Ara | 5.25 | B2IIIne | $287^{+23}_{-20}$ | $292^{+15}_{-14}$ | LABOCA | ⋯ | 5 |
| 6510 | $\alpha$ Ara | 2.95 | B2Vne | $82.0^{+6.1}_{-5.3}$ | $239^{+104}_{-56}$ | LABOCA | ATCA | 3 |
| 6712 | 66 Oph | 4.60 | B2Ve | $200^{+11}_{-10}$ | $143^{+11}_{-10}$ | JVLA | MRT, ATCA, VLA | 4 |
| 7318 | 2 Vul | 5.44 | B0.5IV | $373^{+58}_{-44}$ | $515^{+47}_{-40}$ | ⋯ | MRT | 1 |
| 7708 | 28 Cyg | 4.93 | B2.5Ve | $317^{+25}_{-22}$ | $188.9^{+8.4}_{-7.7}$ | ⋯ | MRT, VLA | 5 |
| 7807 | ⋯ | 5.90 | B2Ven | $415^{+55}_{-43}$ | $377^{+12}_{-11}$ | ⋯ | MRT | 5 |
| 8047 | 59 Cyg | 4.75 | B1ne | $435^{+97}_{-67}$ | $399^{+59}_{-45}$ | ⋯ | VLA | 2 |
| 8053 | 60 Cyg | 5.43 | B1Ve | $467^{+98}_{-69}$ | $402^{+16}_{-15}$ | ⋯ | MRT | 5 |
| 8260 | $\epsilon$ Cap | 4.55 | B2.5Vpe | $324^{+20}_{-18}$ | $153^{+18}_{-15}$ | JVLA | MRT | 5 |
| 8375 | ⋯ | 5.86 | B2.5Ve | $314^{+44}_{-34}$ | $481^{+36}_{-32}$ | ⋯ | VLA | 5 |
| 8402 | $o$ Aqr | 4.69 | B7IVe | $133.5^{+4.2}_{-4.0}$ | $144.5^{+5.6}_{-5.2}$ | JVLA | MRT, VLA | 3 |
| 8539 | $\pi$ Aqr | 4.64 | B1Ve | $240^{+17}_{-15}$ | $286^{+38}_{-30}$ | ⋯ | ATCA | 2 |
| 8731 | EW Lac | 5.43 | B4IIIep | $252^{+19}_{-17}$ | $298^{+13}_{-12}$ | JVLA | JCMT, MRT, VLA | 3 |

Table 1
(Continued)

| HR | Name | V Magnitude | Spectral Type | Hipparcos Distance (pc) | Gaia DR2 Distance (pc) | New Radio Data (2010+) | Archival Radio Data | Figure Set |
|---|---|---|---|---|---|---|---|---|
| 8762 | o And | 3.62 | B6IIIpe+A2p | $211^{+26}_{-21}$ | $108.5^{+10.8}_{-9.0}$ | ... | MRT | 2 |
| 8773 | β Psc | 4.52 | B6Ve | $125.2^{+3.5}_{-3.4}$ | $129.8^{+4.2}_{-4.0}$ | LABOCA | ATCA | 4 |
| 9076 | ε Tuc | 4.47 | B9IV | $114.4^{+2.4}_{-2.3}$ | $107.9^{+3.3}_{-3.1}$ | ... | ATCA | 1 |

Table 2
Spectroscopic Binaries in the Sample

| HR Number | Star | P (days) | e | Type of Companion | References |
|---|---|---|---|---|---|
| 193 | o Cas | 2.8 yr | 0 | MS binary | Koubský et al. (2010) |
| 264 | γ Cas[a] | 203.5 | 0 | ? | Nemravová et al. (2012) |
| 472 | α Eri[b] | 7.07 yr | 0.74 | A-type MS | P. Kervella et al. (2019, in preparation) |
| 496 | φ Per | 126.7 | 0 | sdOB | Mourard et al. (2015) |
| 1910 | ζ Tau | 133.0 | 0 | ? | Ruždjak et al. (2009) |
| 2142 | HR 2142 | 80.9 | 0 | sdOB | Peters et al. (2016) |
| 2845 | β CMi[a] | 170.4 | 0 | sdOB cand. | Dulaney et al. (2017) |
| 3034 | o Pup | 28.9 | 0 | sdOB | Koubský et al. (2012a) |
| 4787 | κ Dra | 61.6 | 0 | ? | Saad et al. (2005) |
| 5953 | δ Sco | 10.8 yr | 0.94 | B-type MS | Miroshnichenko et al. (2013) |
| 8047 | 59 Cyg | 28.2 | 0.14 | sdOB | Peters et al. (2013) |
| 8053 | 60 Cyg | 146.6 | 0 | sdOB | Koubský et al. (2000) |
| 8260 | ε Cap | 128.5 | 0 | ? | Rivinius et al. (2006) |
| 8539 | π Aqr | 84.2 | 0 | ? | Zharikov et al. (2013) |
| 8762 | o And | 33.0 | 0.24 | B-type MS | Zhuchkov et al. (2010) |

**Notes.**
[a] SED previously modeled by Klement et al. (2017b).
[b] Included among wider binaries (Section 3.1.2).

The sample contains several stars for which previous modeling of the SED structure at radio wavelengths implies disk truncation by close companions (Klement et al. 2015, 2017b). Among this subsample, for γ Cas, the companion has long been known, and for β CMi, the companion has been detected recently (Dulaney et al. 2017), although the detection was disputed by Harmanec et al. (2019). The suspected close companion remains undetected by any other method for η Tau, EW Lac, ψ Per, and β Mon A.

In addition to the stars in Table 2, there are also a number of suspected spectroscopic binaries, whose status has not been confirmed. One of these is χ Oph, which was tentatively proposed as a single-lined spectroscopic binary by Abt & Levy (1978) and Harmanec (1987) but not confirmed with interferometric observations by Tycner et al. (2008). The stars 28 Cyg, HR 7807, and ι Ara were reported as candidate Be+sdOB binaries, and HR 2249 was reported as a potential candidate (Wang et al. 2018). Finally, the sample stars 25 Ori, p Car, η Cen, θ Cir, ζ Oph, 66 Oph, and 2 Vul were reported to show RV variations possibly caused by binarity (Chini et al. 2012). Short-period RV variability was also reported for 17 Tau (4.3 days), η Tau (4.1 days), and 48 Per (16.6 days) by Jarad et al. (1989). However, these results have not been verified and may be due to different reasons than orbital motion.

*3.1.2. Wide Binaries*

There are three wider binaries in our sample for which the binary nature is relevant to the SED analysis: the eccentric binary α Eri, the triple system β Mon, and the interferometric binary δ Cen. These are discussed individually in detail in Appendix A.2. Several of the spectroscopic binaries discussed above have also been reported to have wide, faint companions: γ Cas (separation ∼2″.1), HR 2142 (∼0″.6), 59 Cyg (∼0″.2), and o And (∼0″.27). Others with reported companions with separations ≲10″ include μ Cen, θ Cir, μ Lup, 66 Oph, and HR 8375, but the companions are also much fainter than the primary CBes. The separation measurements were taken from Mason et al. (1997, 2009), Oudmaijer & Parr (2010), and Tokovinin et al. (2014).

## 4. Observations

The primary data set analyzed in this work consists of all available radio (submillimeter to centimeter) measurements of CBes, including new, previously unpublished measurements from the NSF's Karl G. Jansky Very Large Array (JVLA); the Large Apex BOlometer CAmera (LABOCA), installed at the Atacama Pathfinder EXperiment[13] (APEX); and the Combined Array for Research in Millimeter-wave Astronomy (CARMA). Our new data include the longest-wavelength detections of CBe stars to date: $\lambda \approx 10$ cm ($\nu \approx 3$ GHz) detections of β CMi and β Mon A.

The wavelength coverage of the radio data ranges from 870 μm to 10 cm, or frequencies from 350 to 3 GHz. The total number of CBes with existing radio flux density measurements is 57, with 24 having been detected in radio on at least one occasion. Nine stars were detected in multiple bands across the centimeter region, allowing for studying the shape of the radio SED in great detail. For the remainder of the targets, we make use of $3\sigma$ level upper limits on flux density at one or more wavelengths.

---

[13] This publication is based on data acquired with the Atacama Pathfinder Experiment (APEX). APEX is a collaboration between the Max-Planck-Institut fur Radioastronomie, the European Southern Observatory, and the Onsala Space Observatory.

Table 3
Details of the IR and Radio Observations

| Mission/Telescope | λ | Epoch of Observation | References |
|---|---|---|---|
| *IRAS* | 12, 25, 60, 100 μm | 1983 Jan–Nov | Helou & Walker ([1988](#)) |
| VLA | 6 cm | 1986 Sep | Taylor et al. ([1987](#)) |
|  | 2 cm | 1987 Dec–1988 Sep | Taylor et al. ([1990](#)) |
|  | 2 cm | 1988 Feb | Apparao et al. ([1990](#)) |
|  | 2, 3.6, 6 cm | 1990 Feb | Dougherty et al. ([1991](#)) |
|  | 2 cm | 1991 Aug | Dougherty & Taylor ([1992](#)) |
| JCMT | 0.8, 1.1 mm | 1988 Aug | Waters et al. ([1989](#)) |
|  | 0.8, 1.1 mm | 1989 Aug | Waters et al. ([1991](#)) |
| IRAM 30 m | 1.2 mm | 1987 Dec–1988 Jan, 1989 Mar | Altenhoff et al. ([1994](#)) |
|  | 1.2 mm | 1991 Feb, 1992 Dec | Wendker et al. ([2000](#)) |
| ATCA | 3.5, 6.3 cm | 1997 Apr–May | Clark et al. ([1998](#)) |
| *AKARI* | 9, 18, 65, 90, 140, 160 μm | 2006 May–2007 Aug | Ishihara et al. ([2010](#)) |
| *WISE* | 3.4, 4.6, 11.6, 22.1 μm | 2010 Jan–Nov | Cutri et al. ([2013](#)) |
| JVLA | 0.7, 1.3, 3.5, 6 cm | 2010 Oct | Klement et al. ([2017b](#)), this work |
|  | 1.3, 3.5, 5, 10 cm | 2017 Feb–May | This work |
| CARMA | 3.09 mm | 2011 May–Oct | Štefl et al. ([2012](#)) |
|  | 1.24, 3.2 mm | 2013 Mar–Jun | Klement et al. ([2017a](#)), this work |
| LABOCA | 0.87 mm | 2011 Apr–Aug | Štefl et al. ([2012](#)) |
|  | 0.87 mm | 2013 Sep | Silaj et al. ([2016](#)), Klement et al. ([2017a](#)), Klement et al. ([2017b](#)), this work |
|  | 0.87 mm | 2015 Jul–Aug | Klement et al. ([2017b](#)), this work |
|  | 0.87 mm | 2015 Nov–2016 Apr | This work |
|  | 0.87 mm | 2017 Aug–Sep, 2018 Apr | This work |

The radio data set was complemented with mid-IR measurements from all-sky space-based surveys, including the *Infrared Astronomical Satellite* (*IRAS*), *AKARI*, and the *Wide-field Infrared Survey Explorer* (*WISE*), mostly covering the interval of 9–60 μm. All of the stars in the sample were successfully detected by at least one of the space missions, with the vast majority detected by two or all three of them. Upper limits for the IR portion of the data set were discarded. The monochromatic fluxes extracted from the mission catalogs had to be color-corrected because of the large bandwidths of the individual filters. This was done in the same way as described in Section 3 of Vieira et al. ([2017](#)). To complement the SED, we also include photometry in the visible region from the catalog of Ducati ([2002](#)).

The data can be roughly divided into two parts: the older data set spanning 1983–1992, which contains *IRAS* measurements in the IR and previously published measurements in the radio, and the newer data set spanning 2006–2017, which includes IR data from *AKARI* and *WISE* and new or recent radio data. The epochs of the data sets used in this study are summarized in Table 3. The archival and previously published data are described in detail in Appendix [B](#).

### 4.1. Radio Submillimeter/Millimeter Flux Density Measurements

#### 4.1.1. APEX/LABOCA

APEX is a modified prototype 12 m antenna for the Atacama Large Millimeter/submillimeter Array (ALMA), located on the Chajnantor Plateau in Chile. LABOCA is a 295 pixel bolometer array camera operating at 870 μm (345 GHz) with a bandwidth of 150 μm (60 GHz) and a half-power beamwidth of $\sim 19\rlap{.}{''}2$. The primary calibrators used to set the flux density scale are the planets Mars, Uranus, and Neptune, and the overall calibration accuracy is $\sim 10\%$ (Siringo et al. [2009](#)).

The older part of the data set was acquired using the wobbler on–off mode, in which the telescope used a chopping secondary mirror (wobbler) to alternate between the observed target and a portion of blank sky nearby at a frequency exceeding the sky noise variations. The other part of the data was taken using the raster map in spiral mode, which does not require a wobbler, since the different pixels (corresponding to individual bolometers in the array) simultaneously scan different areas on the sky. This method produces a fully sampled map with homogeneous noise for an area of about $8'$, and, while requiring longer observing times than the wobbler mode, it should provide more reliable results (Siringo et al. [2009](#)). For more details about the technical description and the observing and calibration procedures, we refer the reader to the official APEX/LABOCA website[14] and Siringo et al. ([2009](#)).

The CRUSH[15] data reduction software (Kovács [2008](#)) was used to process the previously unpublished data and reprocess the data for 48 Lib published by Silaj et al. ([2016](#)). To ensure proper flux calibration of the science scans, we inspected the flux calibrator scans and manually adjusted the flux scaling factors—computed as the ratio between the true (expected) and measured flux of the flux calibrator—for the reduction of the science target scans. The residual rms map noise in the vicinity of the stellar position was adopted as the $1\sigma$ flux measurement uncertainty. In cases of nondetection, we use $3\sigma$ upper limits. We adopted the published flux density values for $\beta$ CMi (Klement et al. [2017a](#)), $\beta$ Mon A (Klement et al. [2017b](#)), and

---
[14] http://www.apex-telescope.org/bolometer/laboca/
[15] http://www.sigmyne.com/crush/index.html

**Table 4**
LABOCA Observations at 0.87 mm

| Star | Flux Density (mJy) | S/N | Date | Obs. Method |
|---|---|---|---|---|
| 48 Lib | 23.5 ± 5.1 | 4.6 | 2013 Sep 20–21 | W |
| p Car | <32.1 | ... | 2017 Aug 27–Sep 7 | R |
| α Ara | 29.2 ± 3.9 | 7.5 | 2013 Sep 21 | W |
| α Col | 62.5 ± 2.9 | 21.6 | 2013 Sep 21 | W |
| α Eri | 65.6 ± 6.9 | 9.5 | 2015 Nov 5 | R |
|  | 46.4 ± 6.6 | 7.0 | 2015 Nov 6 | R |
|  | 72.2 ± 6.3 | 11.5 | 2015 Nov 7 | R |
|  | 57.4 ± 8.1 | 7.1 | 2015 Nov 8 | R |
|  | 65.3 ± 7.1 | 9.2 | 2015 Nov 9 | R |
|  | 72.4 ± 18.7 | 3.9 | 2016 Apr 8 | R |
| β CMi | 38.6 ± 3.1 | 12.5 | 2013 Sep 23 | W |
| β Mon A | 23.5 ± 3.5 | 6.7 | 2015 Jul 30–Aug 26 | W |
| β Psc | 22.0 ± 5.8 | 3.8 | 2017 Sep 7 | R |
| δ Cen | 129.2 ± 9.5 | 13.6 | 2017 Aug 27 | R |
| δ Sco | 69.5 ± 4.2 | 11.2 | 2011 Apr 25 | W |
|  | 78.4 ± 6.8 | 16.5 | 2011 Apr 25 | W |
|  | 42.0 ± 3.7 | 11.0 | 2011 Jul 28 | W |
|  | 51.4 ± 6.7 | 7.7 | 2011 Aug 2 | W |
| ζ Tau | <10.5 | ... | 2013 Sep 20–23 | W |
| η Cen | 23.4 ± 4.3 | 5.4 | 2013 Sep 20 | W |
|  | 25.2 ± 3.0 | 8.4 | 2015 Jul 26 | W |
| ι Ara | <18.6 | ... | 2017 Aug 27 | R |
| κ CMa | <16.8 | ... | 2018 Apr 17–19 | R |
| o Pup | <20 | ... | 2015 Aug 29 | W |
| χ Oph | 71.5 ± 6.7 | 10.7 | 2017 Aug 27 | R |
| ω CMa | 22.9 ± 4.3 | 5.3 | 2015 Aug 29 | W |
|  | <15.0 | ... | 2017 Sep 6–7 | R |
| ω Ori | <15.0 | ... | 2017 Sep 6 | R |

**Note.** W—wobbler on–off; R—raster map in spiral mode.

δ Sco (Štefl et al. 2012). All LABOCA measurements, including the previously published ones, are summarized in Table 4.

### 4.1.2. CARMA

CARMA was a telescope array located in the Inyo Mountains of California, consisting of 23 antennas of three different sizes—six 10.4 m, nine 6.1 m, and eight 3.5 m—operating at millimeter wavelengths.

The 10 and 6 m subarray in the D configuration (baseline of 11–150 m) was used to secure the observations for δ Sco (published by Štefl et al. 2012), β CMi (published by Klement et al. 2015), and ζ Tau. The center frequencies of the observations were 245 and 100 GHz with bandwidths of 8 GHz, which corresponds to central wavelengths of ∼1 and ∼3 mm, respectively. The angular resolution was ∼2″ and ∼5″ for the two bands, respectively. The published values were adopted for β CMi and δ Sco. The previously unpublished data for ζ Tau were reduced using the Miriad[16] software package and the same standard reduction procedure for wide-band continuum data detailed in the Miriad CARMA Cookbook.[17] The Miriad tasks *mossdi* and *restor* were used for the final imaging. The peak flux of a Gaussian fit to the stellar emission and the residual rms in the vicinity of the stellar position were used as the final flux densities and errors, respectively. For the analysis, we also included a 10% flux calibration error to account for the uncertainty in the assumed fluxes of the flux calibrators. The measurements are summarized in Table 5.

### 4.2. Radio Centimeter Flux Density Measurements

#### 4.2.1. JVLA

For a detailed description of the 2010 JVLA data for η Tau, EW Lac, γ Cas, and ψ Per, we refer the reader to Section 3.3.3 of Klement et al. (2017b). During the same observing run and using the same observing strategy, data were also obtained for 66 Oph, ε Cap, o Aqr, ζ Tau, and 48 Lib. Corresponding to central wavelengths of 0.7, 1.3, 3.5, and 6.0 cm, 66 Oph, ε Cap, and o Aqr were observed in the Q, K, X, and C bands, respectively. Also, ζ Tau was successfully observed only in the K and C bands, while 48 Lib was observed only in the C band.

The previously unpublished 2010 data were reduced using the Common Astronomy Software Applications (CASA) reduction software version 5.4.0-68 following the standard procedure for wide-band continuum data,[18] with special care given to proper flagging of parts of the data affected by technical or atmospheric issues. The CASA task *clean* with robust +0.5 weighting of the visibilities was used to produce the final images. The peak flux densities of Gaussian fits to the stellar emission were taken to represent the final flux densities of the unresolved stellar sources, and the rms noise in the vicinity of the stellar positions was used as the 1σ measurement uncertainties. In cases of nondetection, we used the image rms to calculate 3σ upper limits. For the analysis, we also assumed an extra flux calibration error of 10% in the Q and K bands and 5% in the X and C bands, following the official VLA observational status summary.[19] The previously unpublished 2010 data are summarized in the upper part of Table 6.

In 2017, we used the JVLA to observe the targets β CMi and β Mon in the K, X, C, and S bands, which corresponded to central wavelengths of 1.36, 3, 5, and 10 cm, respectively. The WIDAR correlator was set up to use the maximum possible frequency range in order to allow for maximum sensitivity. The frequency range of the respective bands was 18–26.5, 8–12, 4–8, and 2–4 GHz. The most compact D-configuration (maximum baseline of 1.03 km) was used for the K- and X-band observations, while the second most compact C-configuration (maximum baseline of 3.4 km) was used for the C and S bands. The average angular resolution (synthesized beamwidth) was 3″.1, 7″.2, 3″.5, and 7″.0 for the bands K, X, C, and S, respectively, and our sources remained unresolved. The CASA software version 5.1.2–4 was used for data processing and calibration, and the same procedure as described above for the 2010 data was used for the final imaging and flux density estimation. For the analysis, we included an extra flux calibration error of 10% for the K band and 5% for the X, C, and S bands. The 2017 JVLA data are summarized in the lower part of Table 6.

The targets β CMi and β Mon A were detected in the K, X, and C bands and, for the first time, in the S band. This constitutes the first-ever detections of a CBe at wavelengths longer than 6 cm. As for the two companions of β Mon A, the component C was detected in the K, X, and C bands, and an upper limit was obtained for the S band. The B component was not detected in any band, and only upper limits could be obtained.

---

[16] http://carma.astro.umd.edu/miriad/
[17] http://carma.astro.umd.edu/miriad/carmacookbook.pdf
[18] https://casaguides.nrao.edu/index.php?title=VLA_Continuum_Tutorial_3C391-CASA5.0.0
[19] https://science.nrao.edu/facilities/vla/docs/manuals/oss/performance/fdscale

Table 5
CARMA Observations

| Star | $\nu$ (GHz) | $\lambda$ (mm) | Flux Density (mJy) | S/N | Date | Flux Cal. | Bandpass Cal. | Phase Cal. |
|---|---|---|---|---|---|---|---|---|
| $\beta$ CMi | 92 | 3.26 | $9.6 \pm 0.6$ | 16.0 | 2013 Jun 9 | Mars | OJ 287 | 0750+125 |
| $\delta$ Sco | 97 | 3.09 | $25.6 \pm 2.0$ | 12.8 | 2011 Apr 3 | MWC 349 | MWC 349 | 1625−254 |
|  | 97 | 3.09 | $13.7 \pm 0.8$ | 17.1 | 2011 Aug 31 | 3C 279 | 3C 279 | 1625−254 |
|  | 97 | 3.09 | $16.4 \pm 1.2$ | 13.7 | 2011 Sep 2 | 1625−254 | 1625−254 | 1625−254 |
| $\zeta$ Tau | 242 | 1.24 | <7.2 | ⋯ | 2013 Mar 23 | Mars | 3C 111 | 0532+075 |
|  | 92 | 3.26 | $2.4 \pm 0.6$ | 4.0 | 2013 Mar 29 | Mars | 0510+180 | 0532+075 |

Table 6
New JVLA Observations

| Star | Band | $\lambda$ (cm) | Flux Density (mJy) | S/N | Date | Flux/Bandpass Calibrator | Phase Calibrator |
|---|---|---|---|---|---|---|---|
| 48 Lib | C | 6.05 | <0.150 | ⋯ | 2010 Oct 31 | 3C 286 | J1543+0757 |
| 66 Oph | Q | 0.69 | <1.32 | ⋯ | 2010 Nov 5 | 3C 286 | J1804+0101 |
|  | K | 1.34 | $0.640 \pm 0.12$ | 5.3 | 2010 Nov 5 | 3C 286 | J1804+0101 |
|  | X | 3.54 | $0.236 \pm 0.030$ | 7.9 | 2010 Nov 5 | 3C 286 | J1804+0101 |
|  | C | 6.05 | $0.091 \pm 0.012$ | 7.6 | 2010 Oct 21 | 3C 286 | J1804+0101 |
| $\epsilon$ Cap | Q | 0.69 | <0.96 | ⋯ | 2010 Oct 28 | 3C 48 | J2129+1538 |
|  | K | 1.34 | <0.231 | ⋯ | 2010 Oct 29 | 3C 48 | J2129+1538 |
|  | X | 3.54 | <0.084 | ⋯ | 2010 Oct 29 | 3C 48 | J2129+1538 |
|  | C | 6.05 | <0.033 | ⋯ | 2010 Oct 29 | 3C 48 | J2129+1538 |
| $\zeta$ Tau | X | 3.54 | <0.140 | ⋯ | 2010 Oct 25 | 3C 48 | J0559+2353 |
|  | C | 6.05 | $0.111 \pm 0.030$ | 3.7 | 2010 Oct 19 | 3C 48 | J0559+2353 |
| $o$ Aqr | Q | 0.69 | <1.05 | ⋯ | 2010 Oct 30 | 3C 48 | J2218+0335 |
|  | K | 1.34 | $0.377 \pm 0.083$ | 4.5 | 2010 Oct 30 | 3C 48 | J2218+0335 |
|  | X | 3.54 | $0.152 \pm 0.028$ | 5.4 | 2010 Oct 30 | 3C 48 | J2218+0335 |
|  | C | 6.05 | $0.0751 \pm 0.0115$ | 6.5 | 2010 Oct 30 | 3C 48 | J2218+0335 |
| $\beta$ CMi | K | 1.36 | $1.388 \pm 0.032$ | 43.4 | 2017 Mar 15 | 3C 286 | J0745+1011 |
|  | X | 3.00 | $0.467 \pm 0.017$ | 27.5 | 2017 Mar 15 | 3C 286 | J0745+1011 |
|  | C | 5.00 | $0.242 \pm 0.010$ | 24.2 | 2017 May 23 | 3C 286 | J0745+1011 |
|  | S | 10.00 | $0.076 \pm 0.016$ | 4.8 | 2017 May 23 | 3C 286 | J0745+1011 |
| $\beta$ Mon A | K | 1.36 | $0.682 \pm 0.038$ | 17.9 | 2017 Feb 17 | 3C 48 | J0607+0834 |
|  | X | 3.00 | $0.294 \pm 0.018$ | 16.3 | 2017 Feb 17 | 3C 48 | J0607+0834 |
|  | C | 5.00 | $0.162 \pm 0.011$ | 14.7 | 2017 May 21 | 3C 48 | J0607+0834 |
|  | S | 10.00 | $0.080 \pm 0.017$ | 5.3 | 2017 May 21 | 3C 48 | J0607+0834 |
| $\beta$ Mon B | K | 1.36 | <0.11 | ⋯ | 2017 Feb 17 | 3C 48 | J0607+0834 |
|  | X | 3.00 | <0.054 | ⋯ | 2017 Feb 17 | 3C 48 | J0607+0834 |
|  | C | 5.00 | <0.033 | ⋯ | 2017 May 21 | 3C 48 | J0607+0834 |
|  | S | 10.00 | <0.051 | ⋯ | 2017 May 21 | 3C 48 | J0607+0834 |
| $\beta$ Mon C | K | 1.36 | $0.444 \pm 0.038$ | 11.7 | 2017 Feb 17 | 3C 48 | J0607+0834 |
|  | X | 3.00 | $0.176 \pm 0.018$ | 9.8 | 2017 Feb 17 | 3C 48 | J0607+0834 |
|  | C | 5.00 | $0.076 \pm 0.011$ | 6.9 | 2017 May 21 | 3C 48 | J0607+0834 |
|  | S | 10.00 | <0.051 | ⋯ | 2017 May 21 | 3C 48 | J0607+0834 |

### 4.3. Spectroscopy

We used Hα line profiles from spectra available in the Be Star Spectra (BeSS) database (Neiner et al. 2011; Neiner 2018) and Be Stars Observation Survey (BeSOS) catalog (Arcos et al. 2018) to investigate the history of disk presence for the sample CBes. The BeSS database assembles professional and amateur spectra from various instruments, while the BeSOS catalog contains high-resolution spectra obtained using the echelle spectrograph PUCHEROS with a resolution of 17,000 in the wavelength range of 420–730 nm (Vanzi et al. 2012). From the BeSS database, we only used spectra with a resolution of at least 10,000. We also make use of earlier Hα line profiles that are available in the published literature (Andrillat & Fehrenbach 1982; Hanuschik et al. 1996; Banerjee et al. 2000).

## 5. SED Analysis

As explained in Section 2.2, a power law in the form $F_\nu \propto \lambda^{-\kappa}$ is an appropriate approximation for the spectral shape of typical CBes from near-IR wavelengths longward, with $\kappa$ being a known function of the density slope $n$, disk flaring exponent $\beta$, and free–free and bound–free Gaunt factors. The convention of using the $F_\nu(\lambda)$ dependence was adopted following Waters et al. (1991) and associated studies. To determine the density slope $n$ from the spectral slopes via Equation (1), we have to specify the temperature- and wavelength-dependent Gaunt factor expressions $g$ and $b$. We

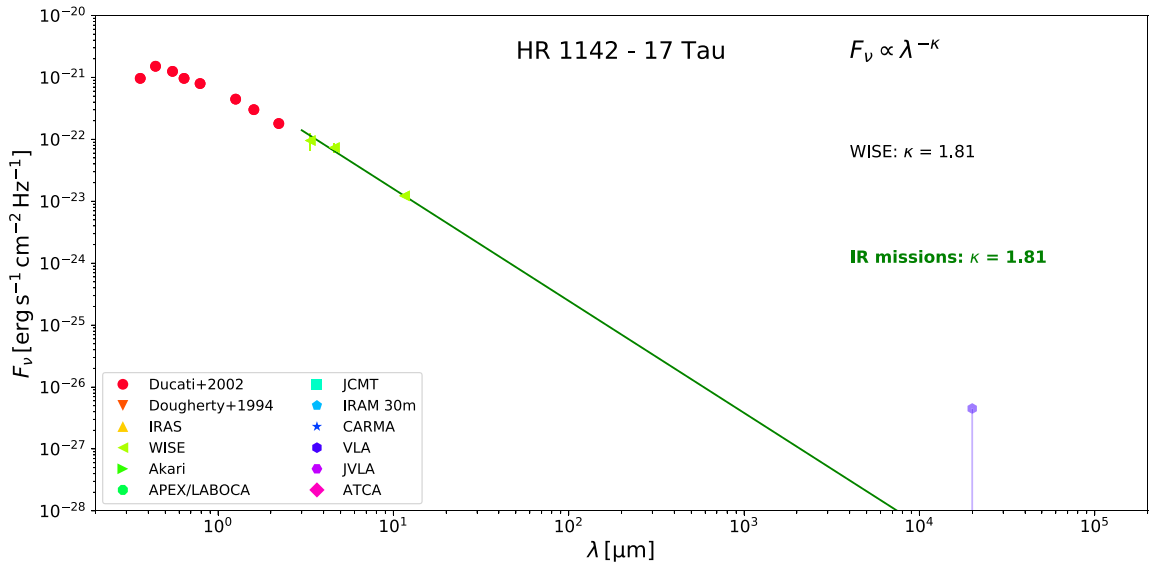

**Figure 1.** SED of 17 Tau. The dashed lines show power-law fits to the data from the IR missions or radio telescopes of the corresponding color. The solid dark green and purple lines show the power-law fits to the entire IR data set and the radio detections in the centimeter region, respectively. The $3\sigma$ upper limits are indicated by a downward-pointing arrow. All five images are available in the Figure Set.

(The complete figure set (5 images) is available.)

use 15 $\mu$m as the representative IR wavelength and 60% of the effective temperature of the central stars derived from their spectral types (according to Cox 2000) as the representative disk temperature for each target. We then use Table A1 of Vieira et al. (2015) to select the Gaunt factor parameters that are closest to the calculated disk temperature for each star. Finally, we assume the isothermal value of 1.5 for the disk flaring parameter $\beta$, as even for nonisothermal disks, an isothermal temperature profile is expected for their outer parts (see Figure 4 of Carciofi & Bjorkman 2008).

In order to search for a turndown in the observed SED, we fit the power law through the IR data and extrapolate it to the radio region. We fit the data using the Levenberg–Marquardt algorithm for nonlinear least-squares curve fitting (e.g., Moré 1978), as implemented in the community Python package for astronomy Astropy version 3.2.1, with weights equal to the inverse of measurement errors. When two or more radio detections in the centimeter region are available, we fit a power law through the centimeter data set and compare the spectral slope to the one in the IR. In case of fewer than two radio detections, we compare the IR slope extrapolation with the available radio data, which may include upper limits only. The SED turndown is established either by the $3\sigma$ upper limits lying an additional $1\sigma$ below the IR extrapolation or by the detections lying below the IR extrapolation by at least $3\sigma$.

The data used in this study span ∼34 yr, which is a timescale over which CBe disks can completely dissipate, reform, or oscillate between the two states. While for stable disks, it is straightforward to establish whether SED turndown is present or not, for the more active stars, we must treat the SED data sets separately and compare contemporaneous measurements only. We do this by making a fit to each individual mission or telescope, provided that at least two detections are available.

To help with assessing the variability of the CBes showing SED turndown and the few special cases, we use historic spectroscopic measurements of the strongest emission line, H$\alpha$, which can provide us with important information about the disk behavior during and between the continuum measurements. Also, since the emission from H$\alpha$ and the IR continuum excess originates from disk regions of similar sizes (Carciofi 2011), the observed line profiles are particularly relevant to the interpretation of the IR observations.

The SED analysis allowed us to separate the studied CBes into several groups, which will be discussed individually below. We start with CBes whose IR measurements indicate little to no disk presence (Section 5.1) and CBes with inconclusive data (Section 5.2), and we follow with stable or weakly variable CBes that show an SED turndown (Section 5.3), CBes exhibiting stronger disk variations (Section 5.4), and three special cases (Section 5.5).

### 5.1. Stars with Little to No Disk Presence during IR Measurements

Out of the entire data set, seven targets were observed in IR when no disk or only a tenuous disk was present, which is revealed by the lack of IR continuum excess. In this group, we include stars with the IR spectral slope exponent $\kappa \geqslant 1.7$. The radio data for five of these stars—17 Tau, 69 Ori, $\lambda$ Cru, 2 Vul, and $\epsilon$ Tuc—are comprised of only upper limits that lie above the extrapolation of the IR spectral slope and are therefore of no further interest in the context of this study. An example SED for these stars is plotted in Figure 1, with the complete figure set (5 figures) available in the online journal.

The other two targets with no disk signature in the IR—$\alpha$ Eri and $\zeta$ Oph—have radio data indicating the presence of a disk at a different time. These two cases will be discussed in more detail in Sections 5.5 and 5.4, respectively.

### 5.2. Stars with Inconclusive SED Measurements

This group includes 11 targets for which a scarce amount of SED data and/or insufficient sensitivity of radio measurements preclude definitive conclusions about the presence of SED turndown.

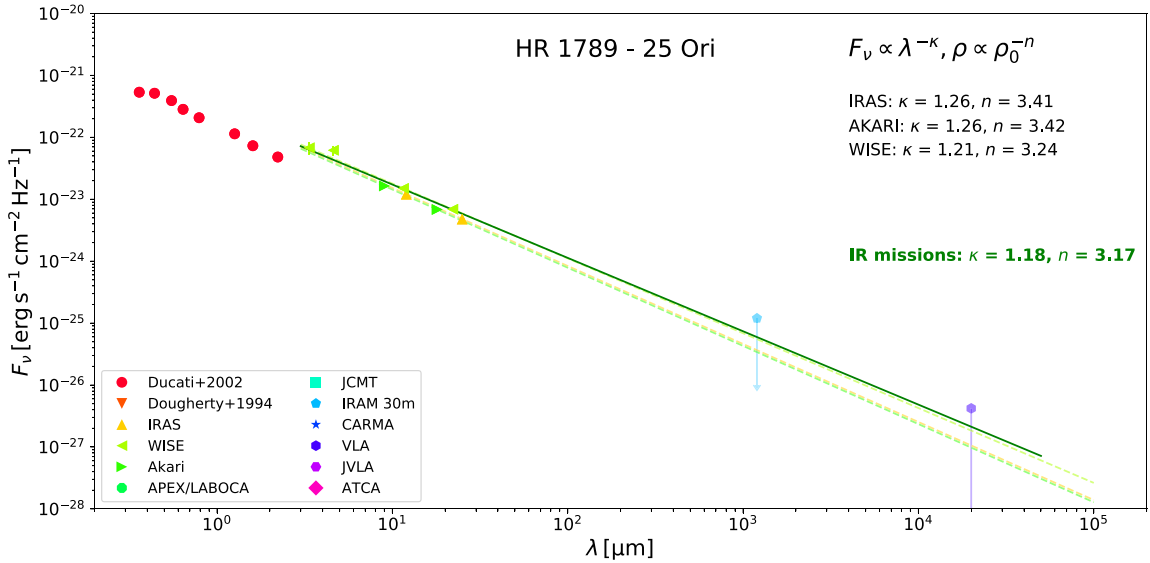

**Figure 2.** Same as Figure 1 but for 25 Ori. All 11 images are available in the Figure Set.
(The complete figure set (11 images) is available.)

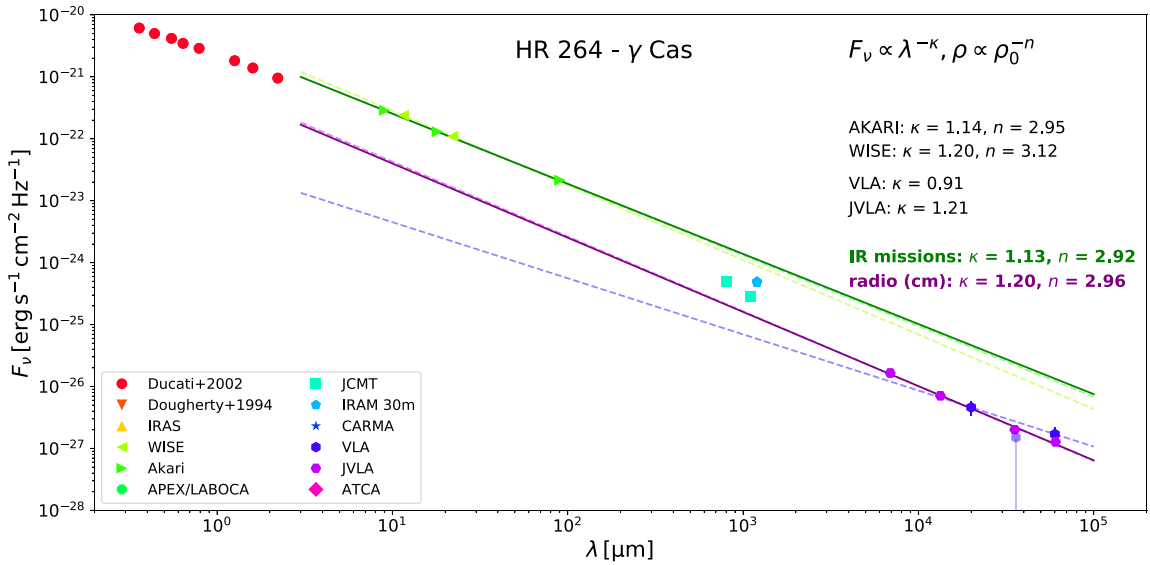

**Figure 3.** Same as Figure 1 but for γ Cas. All 20 images are available in the Figure Set.
(The complete figure set (20 images) is available.)

Four of the stars in this group are known to have close binary companions, which are likely to cause strong SED turndown that we are unable to detect with the given sensitivity of the radio measurements. These include π Aqr, o And, and the Be+sdOB binaries o Pup and 59 Cyg.

The other seven CBes in this group, namely, 25 Ori, ω Ori, HR 2249, β Mon B, HR 2545, θ Cir, and μ Lup, have not been confirmed to have close companions, although HR 2249 is a potential candidate Be+sdOB binary (Wang et al. 2018), and 25 Ori and θ Cir are unconfirmed spectroscopic binaries (Chini et al. 2012). In Section 5.3, β Mon B is discussed in detail, along with the A and C components. Figure 2 shows an example SED from this group, with the complete figure set (11 figures) available in the online journal.

### 5.3. Stars with Stable or Weakly Variable Disks Showing SED Turndown

Twenty stars in the sample have had a stable presence of circumstellar disks over the epochs spanned by the IR and radio measurements. In addition to the IR data being in agreement with each other, we confirm the stability by investigating Hα line profile measurements published in the literature and available online in the BeSS and BeSOS databases. In these cases, we can be confident that the observed SED all the way to the radio regime reflects the physical structure of the disk from its inner to outer parts, and therefore that the observed SED turndown indicates a change in the density structure of the outer parts.

Table 7
Spectral ($\kappa$) and Density ($n$) Slopes of Stable Stars with SED Turndown

| HR Number | Name | $\kappa$ (IRAS) | $\kappa$ (AKARI) | $\kappa$ (WISE) | $\kappa$ (IR) | $n$ (IR) | $\kappa$ (radio cm) | $n$ (radio cm) |
|---|---|---|---|---|---|---|---|---|
| 264 | $\gamma$ Cas[a] | ... | 1.14 | 1.20 | 1.13 | 2.92 | 1.20 | 2.96 |
| 1087 | $\psi$ Per | 0.80 | 0.91 | 0.94 | 0.91 | 2.56 | 1.21 | 2.89 |
| 1165 | $\eta$ Tau | 0.97 | 1.11 | ... | 1.08 | 2.89 | 1.25 | 3.02 |
| 1273 | 48 Per | 0.87 | 0.84 | 0.93 | 0.87 | 2.46 | ... | ... |
| 1622 | 11 Cam | 0.87 | 0.27 | 0.53 | 0.55 | 2.06 | ... | ... |
| 1660 | 105 Tau | 0.87 | 0.59 | 0.69 | 0.66 | 2.17 | ... | ... |
| 1910 | $\zeta$ Tau[a] | 1.16 | 1.09 | 1.24 | 1.13 | 2.98 | ... | ... |
| 1956 | $\alpha$ Col | 1.03 | 1.00 | 1.02 | 0.98 | 2.69 | ... | ... |
| 2142 | HR 2142[a] | 1.03 | 1.00 | 1.12 | 1.09 | 2.85 | ... | ... |
| 2356 | $\beta$ Mon A | 1.15 | 0.94 | ... | 0.95 | 2.59 | 1.11 | 2.68 |
| 2358 | $\beta$ Mon C | 1.15 | 0.94 | ... | 0.95 | 2.59 | 1.32 | 3.26 |
| 2845 | $\beta$ CMi[a] | 1.04 | 1.10 | 1.08 | 1.15 | 3.13 | 1.38 | 3.44 |
| 4140 | p Car | ... | 1.09 | 1.08 | 0.98 | 2.65 | ... | ... |
| 4621 | $\delta$ Cen | ... | 0.90 | 1.08 | 0.87 | 2.44 | ... | ... |
| 4787 | $\kappa$ Dra[a] | 1.10 | 0.90 | 1.03 | 1.03 | 2.78 | ... | ... |
| 5941 | 48 Lib | 1.10 | 0.90 | 1.03 | 0.77 | 2.36 | ... | ... |
| 6118 | $\chi$ Oph | 0.84 | 0.97 | 1.09 | 0.96 | 2.58 | ... | ... |
| 6510 | $\alpha$ Ara | 1.10 | 1.17 | 1.17 | 1.21 | 3.16 | ... | ... |
| 8402 | o Aqr | 0.59 | 0.95 | 1.17 | 0.95 | 2.62 | 1.07 | 2.57 |
| 8731 | EW Lac | 0.61 | 0.82 | 0.87 | 0.85 | 2.44 | 1.04 | 2.54 |

**Note.**
[a] Confirmed close binary.

Table 8
Spectral Slopes of Variable Stars with SED Turndown

| HR Number | Name | $\kappa$ (IRAS) | $\kappa$ (AKARI) | $\kappa$ (WISE) | $\kappa$ (Radio cm) |
|---|---|---|---|---|---|
| 496 | $\phi$ Per[a] | 1.09 | 0.75 | 1.08 | ... |
| 1423 | 228 Eri | 0.79 | 1.17 | 1.26 | ... |
| 2538 | $\kappa$ CMa | 1.01 | 1.20 | 1.09 | ... |
| 2911 | OW Pup | 1.01 | 0.55 | 0.84 | ... |
| 6712 | 66 Oph | 0.82 | 1.30 | 1.52 | 1.29 |
| 8773 | $\beta$ Psc | 0.57 | 1.17 | 0.94 | ... |

**Note.**
[a] Confirmed close binary.

All 20 CBes with overall stable disks and sufficient data coverage clearly show a turndown between the IR and radio region of the SED. This part of the sample includes five known close CBe binaries (see Section 3.1.1 for details) and 15 CBes for which no close companion has been reported. The individual stars are discussed in detail in Appendix C.1. In Table 7, we list the spectral slopes and the estimates of the density slopes $n$ for these 20 stars. An example plot of the SED is shown in Figure 3, with the complete figure set (20 figures) available in the online journal.

### 5.4. Stars with Strong Disk Variations

The sample contains 18 CBes that show clear variability in IR fluxes caused by strongly variable disks. To search for signs of SED turndown in these stars, we compare contemporary data in the IR and radio and examine the history of the disk presence as seen in the varying strength of their H$\alpha$ emission. Overall, we were able to detect strong signs of SED turndown in six of these stars, including one known binary, and these six stars are discussed individually in Appendix C.2.1. The results of the SED fitting are given in Table 8. An example SED plot for the variable CBes with SED turndown is shown in Figure 4, with the complete figure set (6 images) available in the online journal.

For the remaining 12 CBes in this group, we are unable to confidently detect the SED turndown due to the complicated history of variability and/or insufficient data. This latter group includes two confirmed close binaries and two Be+sdOB candidates. The stars in this group are discussed individually in Appendix C.2.2. An example SED plot is shown in Figure 5, with the complete figure set (12 images) available in the online journal.

### 5.5. Special Cases

In this section, we include three CBes that do not fit well into any of the other groups. They are discussed in detail individually in Appendix C.3. Two of these stars—$\delta$ Sco and $\alpha$ Eri—are eccentric binaries in which the companion approaches the primary CBe close enough to affect the disk structure only during periastron passage. These stars were observed at radio wavelengths at times close to the periastron passages specifically to assess the possible disruption of the disk as the companions pass nearby. The presence of an even closer companion in the case of $\delta$ Sco is highly unlikely (Miroshnichenko et al. 2013).

The other special case is $\omega$ CMa, which is well known for quasi-periodic phases of disk growth and dissipation (Ghoreyshi et al. 2018). We observed $\omega$ CMa at the radio wavelength twice in order to detect the changes in its outer disk during the phases of a well-developed disk and after the onset of subsequent phases of disk dissipation. The SED plot for $\omega$ CMa is shown in Figure 6, with the full figure set (3 figures) available in the online journal.

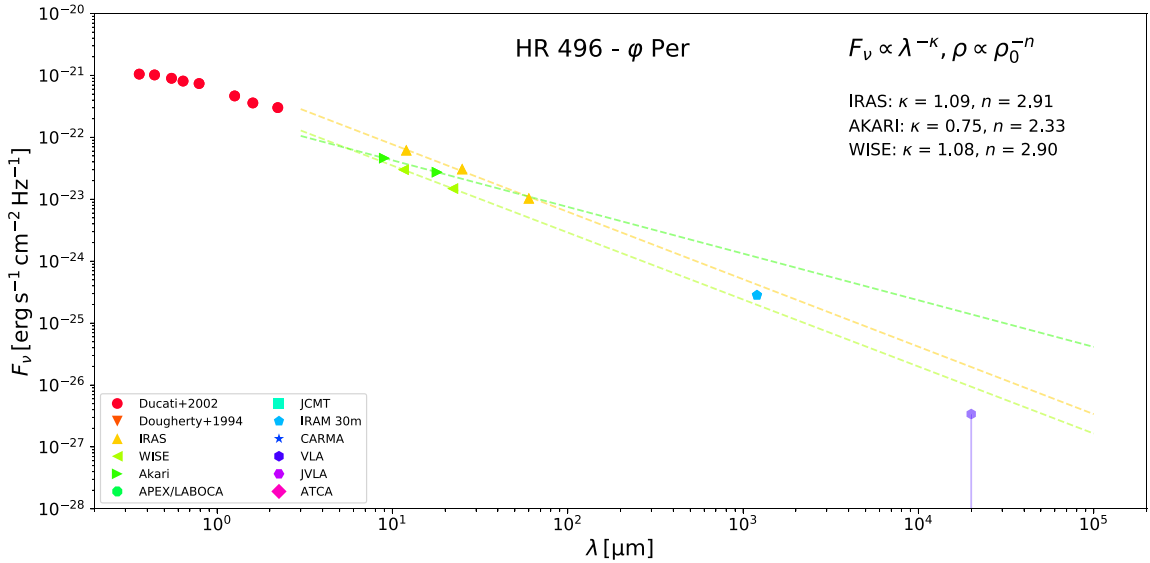

**Figure 4.** Same as Figure 1 but for $\varphi$ Per. All six images are available in the Figure Set.
(The complete figure set (6 images) is available.)

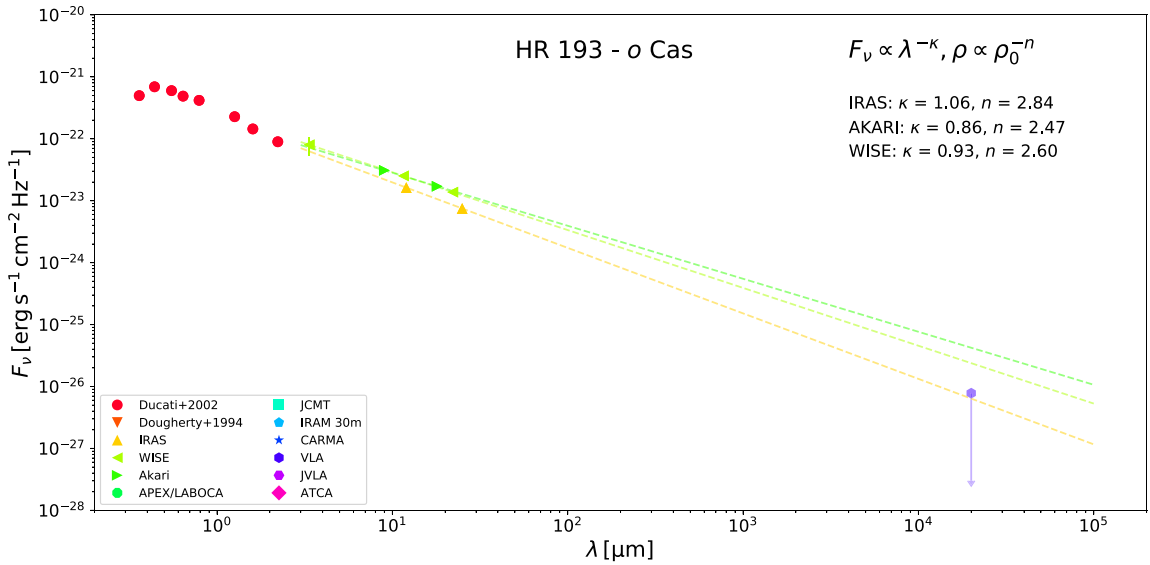

**Figure 5.** Same as Figure 1 but for $o$ Cas. All 12 images are available in the Figure Set.
(The complete figure set (12 images) is available.)

## 6. Summary and Discussion

We analyzed the SEDs of all CBes with available radio data (57 sources total) by fitting a power law through the IR and radio spectral regions. All of the studied CBes with sufficient data coverage and stable disk behavior show an SED turndown between the IR and radio parts of the SED. In the context of our current understanding of CBe disks as VDDs, the presence of an SED turndown implies a change in the density structure at larger distances from the star that likely originates from disk truncation by often unseen close binary companions. The information on the SED fitting for the stars showing SED turndown is summarized in Tables 7 and 8.

For stars with stable and developed disks, the values of the spectral slope exponent $\kappa$ in the IR region range from ∼0.6 to ∼1.2, corresponding to density slope exponents between ∼2.1 and ∼3.1. Histograms of these two quantities are plotted in Figure 7. The distribution of the density exponents is very similar to that derived from a larger sample by Vieira et al. (2017; see their Figure 7).

The spectral slope in the radio centimeter region ranges from 1.0 to 1.4 for nine stars with sufficient (mostly JVLA) data (see Tables 7 and 8). Two of these nine stars—$\gamma$ Cas and $\beta$ CMi— are confirmed close binaries. The distribution of $\kappa$ and $n$ in the radio for the stars with stable disks is shown by dashed lines in Figure 7. While the radio slope is in all cases steeper than the IR slope, it is still shallow with respect to what is expected from sharp truncation, which would result in a spectral slope close to 2.0 longward of the turndown (see Figure 11 of Vieira et al. 2015). Throughout this work, we have used the term truncation to describe any kind of outer disk perturbation, but it is clear that sharp truncation is almost certainly not the case in the majority of close CBe binaries. We interpret the shallow radio slopes as another piece of growing evidence suggesting that CBe disks are circumbinary, with small and often invisible companions unable to fully accrete the outer disk material.

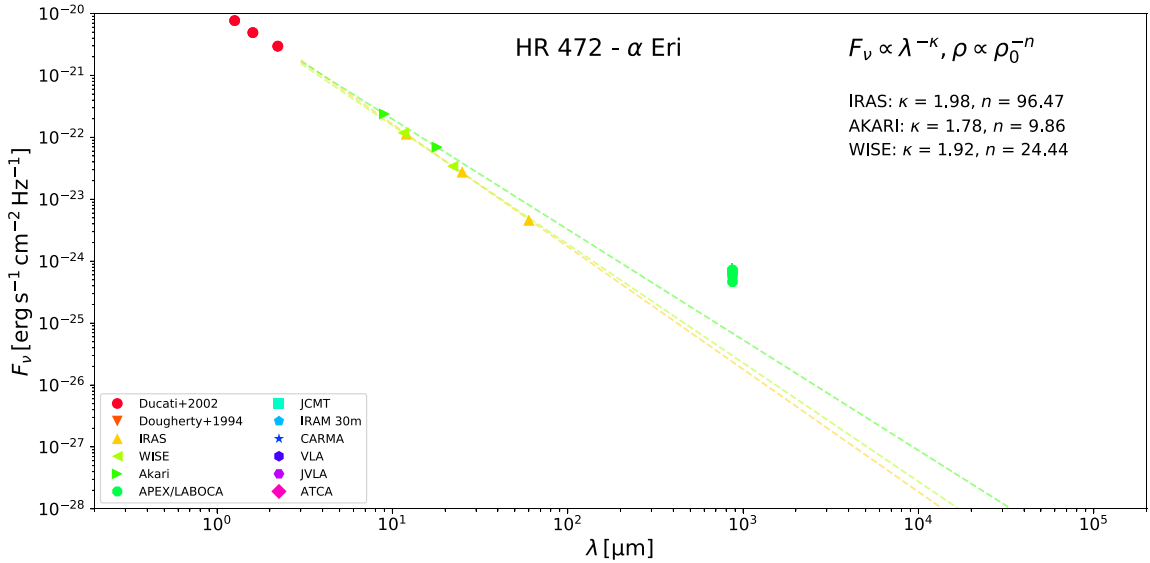

**Figure 6.** SED of $\alpha$ Eri. All three images are available in the Figure Set.
(The complete figure set (3 images) is available.)

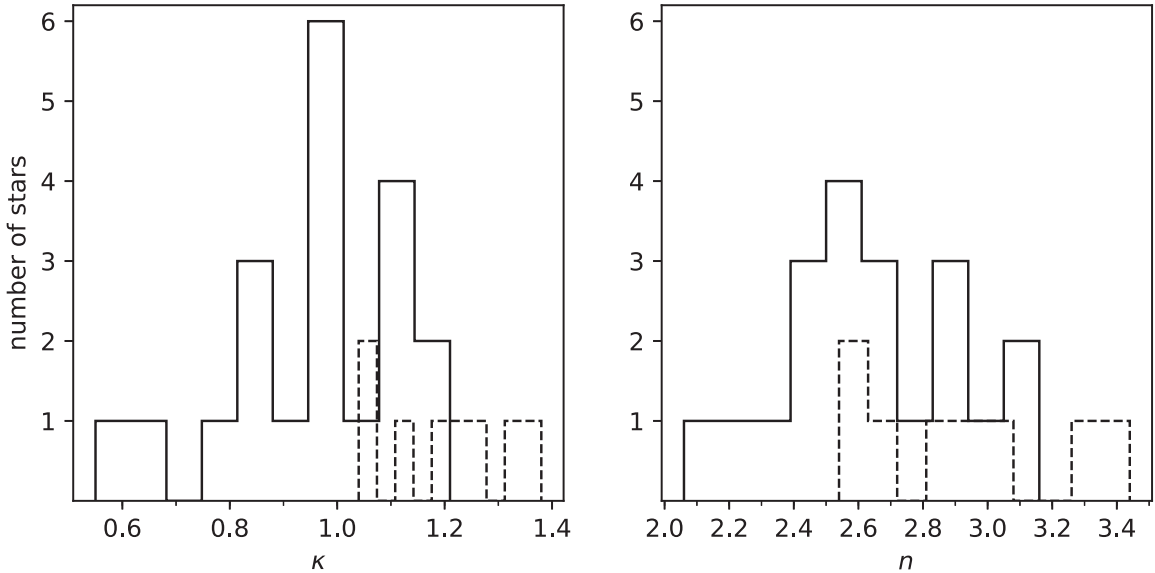

**Figure 7.** Histograms showing the distribution of $\kappa$ (left panel) in the IR (solid line) and radio (dashed line) and $n$ (right panel) in the IR (solid line) and radio (dashed line) among the CBes with stable or weakly variable disks.

The presence of close binary companions that have undergone past mass transfer between the components provides us with the solution to the origin of the rapid rotation of CBes, which is one of their most characteristic properties and a necessary condition leading to the occurrence of outflowing circumstellar disks. Compared to the case of BeXRBs (not represented in our sample), less massive sdOB companions are notoriously hard to detect around bright CBes. However, if they can be found to be prevalent, then most CBes could originate from similar binary progenitors. The SED analysis presented in this study does not allow us to investigate the nature of the suspected companions, and we are unable to distinguish whether they might be sdOB stars or simply late-type MS dwarfs. However, in cases of known close binaries where the nature of the companion has been confirmed, they are sdOB stars (excluding the special cases of $\delta$ Sco and $\alpha$ Eri and the peculiar binary $o$ And). A systematic search for specific signatures of hot sdOB companions in the spectra of CBes showing an SED turndown would be worthwhile in order to assess the nature of the suspected companions.

Besides close binary companions, which definitely truncate CBe disks in at least some cases (see, e.g., the known CBe binaries with SED turndown listed in Appendix C.1.1), it cannot be ruled out that in some cases, the observed SED turndown could be caused by a change in the outer disk density structure due to another reason, such as strong nonisothermal effects. On a purely observational basis, there is a striking similarity between the SEDs of confirmed close binaries, such as $\gamma$ Cas and $\beta$ CMi, and seemingly single stars, such as $\psi$ Per or $\eta$ Tau, which implies that the observed spectral turndown in the latter stars is likely also caused by close binary companions. However, to be able to unambiguously confirm that the observations are compatible with the effects of close companions, realistic hydrodynamical models for both the binary and nonbinary cases; predictions of the corresponding observables, including the radio spectra; and high-quality radio data of a larger number of CBes in the centimeter region would be needed.

## 7. Conclusions

In this work, we analyzed the IR and radio SEDs of 57 bright CBes. We were able to detect an SED turndown for all of the 26 sample stars with sufficient data coverage and developed disks (Tables 7 and 8). Of these, only six are previously confirmed close binaries, with one being a confirmed Be+sdOB system. The other 20 CBes with an SED turndown should now be considered as close binary candidates and potential Be+sdOB candidates.

Although we are unable to confirm the nature of the suspected companions, the fact that we detect signs of invisible companions in all stars with sufficient data provides further observational evidence for the past evolution of mass-transferring binary systems as the cause of the spin-up of the present-day CBes. Furthermore, the radio SED slopes of stars with enough radio data are incompatible with sharp disk truncation, adding another piece to the growing amount of evidence indicating a circumbinary nature for CBe disks. In this case, the typical CBe observables, such as emission lines and IR excess, all originate from the circumprimary part of the disk, with the circumbinary part being observable at long radio wavelengths only. Such a tenuous circumbinary part of the disk could possibly extend to much larger dimensions than previously thought. The measured size of the only CBe disk resolved in the radio—$\psi$ Per—supports this scenario (see Section 5.4 and Figure 5 of Klement et al. 2017b).

For 28 sample stars, we were prevented from reaching definitive conclusions regarding the presence of the SED turndown due to several factors, including scarcity of data, insufficient sensitivity of the flux density measurements, strong variability of the disks, and little to no disk presence during the observations. Four of these stars were previously confirmed to be close binaries, and two more are Be+sdOB candidates. We also described the SEDs and radio flux variations of three special cases, which will be studied in further detail in dedicated publications: the highly eccentric binaries with MS companions $\alpha$ Eri and $\delta$ Sco and the quasi-periodically outbursting CBe $\omega$ CMa.

In this work, we also aimed to assemble the complete radio data set of CBes, which will be valuable for future detailed studies dedicated to modeling of the individual stars. We presented new submillimeter/millimeter measurements of 14 stars from APEX/LABOCA and one star from CARMA, as well as new high-precision multiband measurements of nine stars using the JVLA. Together with JVLA data published by Klement et al. (2017b), nine CBes—66 Oph, EW Lac, $\beta$ CMi, $\beta$ Mon A, $\beta$ Mon C, $\gamma$ Cas, $\eta$ Tau, $o$ Aqr, and $\psi$ Per—now have multiple high-quality detections at centimeter wavelengths, which will allow for detailed modeling of the outer disks. For two of these stars—$\beta$ CMi and $\beta$ Mon A—we were able to get a reliable detection at the wavelength of 10 cm—the longest wavelength at which a CBe was detected so far.

R.K. is grateful for a postdoctoral associateship funded by the Provost's Office of Georgia State University.

A.C.C. acknowledges support from CNPq (grant 307594/2015-7).

D.M.F. acknowledges FAPESP (grant 2016/16844-1).

M.C. is thankful for support from FONDECYT project 1190485 and Centro de Astrofísica de Valparaíso Chile.

This publication is partly based on observations made with ESO telescopes at the La Silla Paranal Observatory under program IDs 092.F-9708, 0.95.F-9709, 296.C-5004, and 099.F-0005.

Support for CARMA construction was derived from the states of California, Illinois, and Maryland; the James S. McDonnell Foundation; the Gordon and Betty Moore Foundation; the Kenneth T. and Eileen L. Norris Foundation; the University of Chicago; the Associates of the California Institute of Technology; and the National Science Foundation.

This publication is partly based on observations made with CARMA under program ID c1100.

The National Radio Astronomy Observatory is a facility of the National Science Foundation operated under cooperative agreement by Associated Universities, Inc.

The VLA data presented here were obtained from programs VLA/10B-143 (AI141) and VLA/18A-348 (AK972).

This publication makes use of data products from the *Wide-field Infrared Survey Explorer*, which is a joint project of the University of California, Los Angeles, and the Jet Propulsion Laboratory/California Institute of Technology, funded by the National Aeronautics and Space Administration.

The James Clerk Maxwell Telescope has historically been operated by the Joint Astronomy Centre on behalf of the Science and Technology Facilities Council of the United Kingdom, the National Research Council of Canada, and the Netherlands Organisation for Scientific Research.

This research has made use of the SIMBAD database, operated at CDS, Strasbourg, France.

This research has made use of the VizieR catalog access tool, CDS, Strasbourg, France.

This work has made use of the BeSS database, operated at LESIA, Observatoire de Meudon, France: http://basebe.obspm.fr.

This work used the BeSOS Catalogue, operated by the Instituto de Física y Astronomía, Universidad de Valparaíso, Chile (http://besos.ifa.uv.cl), and funded by Fondecyt iniciación No. 11130702.

*Software:* CRUSH (Kovács 2008), MIRIAD (Sault et al. 1995), CASA (McMullin et al. 2007), Astropy (Astropy Collaboration et al. 2013), Matplotlib (Hunter 2007).

Appendix A
Known Binaries in the Sample

*A.1. Close Binaries*

*HR 193–o Cas.* This is a peculiar binary in which the A component is the CBe and the secondary is composed of a pair of late-B or early-A dwarfs orbiting each other with a very short period of about 4 days (Grundstrom 2007). The semimajor axis of the primary orbit is 17 mas, and the period is ∼2.8 yr, as derived from both interferometry and RVs (Koubský et al. 2010).

*HR 264—γ Cas.* The prototypical CBe is a single-lined binary with a faint and low-mass companion (Harmanec et al. 2000; Miroshnichenko et al. 2002; Nemravová et al. 2012) whose nature remains disputed (e.g., Smith et al. 2012; Wang et al. 2017). Also, γ Cas is a source of X-rays, which are thought to arise from the interaction of small-scale stellar magnetic fields with the disk close to the star (Smith et al. 2016; Smith 2019) or from accretion onto a degenerate companion (Hamaguchi et al. 2016; Postnov et al. 2017).

*HR 496—φ Per.* This is the first CBe confirmed to have an sdOB companion. It was first noticed due to the He II λ4686 emission line moving in antiphase to the CBe, which can be explained by its origin in the accretion disk around the subdwarf companion (Poeckert 1981; Štefl et al. 2000). The faint, low-mass secondary was detected in UV spectra from *IUE* (Thaller et al. 1995), and the double-line orbit was derived from UV spectra taken with the *HST* (Gies et al. 1998). Hummel & Štefl (2001, 2003) were able to derive that the geometry of the He I emission is compatible with originating from the part of the disk facing the companion, while the Fe II line emission was found to be axisymmetric with respect to the primary CBe. Most recently, Mourard et al. (2015) directly detected the companion by OLBI measurements and updated the orbital solution.

*HR 1910—ζ Tau.* This is a long-known single-line spectroscopic binary CBe (Harmanec 1984; Ruždjak et al. 2009). The nature of the companion has not been confirmed, and the possibilities include a rather cool sdOB or a late-type MS companion (Schaefer et al. 2010). In the past, ζ Tau showed observational signatures of remarkably stable one-armed density waves, which were propagating through the disk with an orbital period of ∼1430 days, but were thought to be unrelated to its binary nature (Carciofi et al. 2009; Štefl et al. 2009).

*HR 2142.* This object has a confirmed sdOB companion. It was noticed as a likely mass-transferring binary for its periodically recurring strong absorption components in the emission lines (so-called "shell" phases; Peters 1976), which were later found to be phase-locked with the orbital period (Peters 1983). Waters et al. (1991) ruled out a cool companion based on the IR SED and suggested that the companion is a hot helium star. The UV spectral signatures of the sdOB companion were detected in *IUE* spectra using cross-correlation techniques by Peters et al. (2016), who also presented evidence for a circumbinary nature of the CBe disk. In their model, the periodic shell line appearance occurs due to gas streams connecting the inner and outer disk parts with the companion inside the disk gap. Occurrence of such variations (also detected in φ Per and HD 55606) seems to require a combination of a high-density disk, a very low mass companion, and an almost edge-on viewing angle. Gas crossing the gap also creates shocks as it hits the denser parts of the disk, which would explain the presence of lines of high-ionization species such as Si IV and C IV. An accretion disk is possibly formed around the companion by the gap-crossing gas, helping to explain the partial obscuration of the companion.

*HR 2845—β CMi.* The spectroscopic binarity of β CMi was suspected by Jarad et al. (1989) and Chini et al. (2012). It was later confirmed by Dulaney et al. (2017), who detected clear orbital variations of Hα emission wings. The violet-to-red peak ratio (V/R) variations seem to be phase-locked with the orbit and therefore are possibly caused by higher-density regions of the disk created in the tidal wake of the orbiting companion (Folsom et al. 2016; Dulaney et al. 2017). Based on the radio SED, Klement et al. (2015) derived a disk size of 35 stellar radii (later revised to 40 by Klement et al. 2017b), which for the orbit of Dulaney et al. (2017) roughly corresponds to a 3:2 orbital resonance ratio between the outer boundary of the disk and the companion. The shape of the radio SED suggests a gradual truncation or circumbinary nature of the disk. Several features seen in high-resolution spectra also suggest the presence of a small, hot, mass-accreting companion. The IR Ca II triplet is in emission, the Ca II λ3934 line is formed in the outer parts of the disk rather than close to the primary, and the C IV λ1548 line, requiring very hot conditions not expected for a B8 star, clearly shows a P Cygni line profile, indicating a radiation-driven stellar wind possibly originating from the vicinity of the hot companion (Klement et al. 2015). The sdOB nature of the companion could not be directly confirmed due to the availability of very few UV spectra (Dulaney et al. 2017).

The binary detection of Dulaney et al. (2017) was recently disputed by Harmanec et al. (2019). The primary criticism of the analysis of the orbit presented by Dulaney et al. (2017) was that the detected RV shifts were caused by seasonal changes in the telluric lines, which might have been the origin of the nearly half-year alias in the period of the orbit. Here we provide added details from the analysis of Dulaney et al. (2017). In that analysis, the strong telluric absorption spectrum was used to both confirm and reestablish the wavelength solution in every observation, and then the telluric lines were removed by means of a template. This should have removed the observational bias mentioned, and the lack of confirmation by Harmanec et al. (2019) could be mostly due to a smaller set of spectroscopic data taken from multiple observatories, most of them having lower spectral resolution than the spectra presented by Dulaney et al. (2017).

*HR 3034–o Pup.* This star is another suspected Be+sdOB system. The evidence comes from the variations of the He I λ6678 line, which indicate that the line is formed in the outer part of the disk, which is illuminated by the hot companion (Rivinius et al. 2012). The RV variations of the line then coincide with the binary orbit, as derived by Koubský et al. (2012a). The nature of the companion could not be directly confirmed, as very few UV spectra exist.

*HR 4787–κ Dra.* This is a single-lined spectroscopic binary showing a phase-locked occurrence of "moving absorption bumps" on top of the emission peaks (Juza et al. 1991; Saad et al. 2005). The exact origin of these variations is unclear. No spectroscopic signature of the companion itself was found in the available spectra, but the system was found to be a weak source of X-rays, indicating that the companion might be a compact object accreting outer disk material (Peters 1982).

*HR 5953–δ Sco*. A short-period companion was suggested by van Hoof et al. (1963; 20 days) and Levato et al. (1987; 83 days), but more recent studies ruled out the presence of such a close companion (Miroshnichenko et al. 2001; Tango et al. 2009). It was shown that δ Sco has a wider companion on a remarkably eccentric orbit in which the component separation decreases from ∼200 mas during apastron to ∼6 mas during periastron (Miroshnichenko et al. 2013). Its orbit was first deduced by Bedding (1993) using OLBI measurements together with previous speckle observations. The full interferometric orbit was derived in detail by Meilland et al. (2011) and Tycner et al. (2011) with an orbital period of ∼10.8 yr. The presence of the eccentric companion was found to not be directly involved in the sudden onset of the primary CBe's disk growth (Miroshnichenko et al. 2001). The direction of the disk rotation, in fact, seems to be strongly misaligned or even opposite with respect to the orbital motion (Che et al. 2012; Štefl et al. 2012). However, the existing disk was found to be perturbed during the last periastron passage, which included a clear decrease of the observed radio fluxes following the periastron (Štefl et al. 2012). The δ Sco binary system is surrounded by a bow shock–like structure and may be a runaway system with an external third component (Miroshnichenko et al. 2013).

*HR 8047–59 Cyg*. A single-lined spectroscopic orbital solution for 59 Cyg was derived by Rivinius & Štefl (2000), who were also the first to suggest that the companion is a faint sdOB star. Subsequently, Maintz et al. (2005) were able to detect the orbital motion of the secondary by the periodic RV shifts of the He II λ4686 line, which originates in the parts of the disk close to the hot companion. Other spectral features, such as the RV of the emission peak of the He I λ6678 line, V/R variations of the same line, and equivalent widths of several other lines, were found to be phase-locked with the binary orbit, which shows a clear influence of the hot companion on the part of the disk facing it. The spectral signature of the sdOB companion was later found in the UV spectra using cross-correlation methods, and its flux contribution in the UV was found to be ∼4% (Peters et al. 2013). It was also recently found that 59 Cyg has a wider, eccentric companion on an ∼906 day orbit (T. Gardner et al. 2019, in preparation), which may help explain the slightly eccentric orbit of the inner sdOB companion.

*HR 8053–60 Cyg*. The signature of an sdOB companion has been directly detected in the observed UV spectra (Wang et al. 2017). The spectroscopic binary orbit was derived from RV variations of the emission wings of the Hα line by Koubský et al. (2000).

*HR 8260–ϵ Cap*. Periodic RV variability of the absorption core of the Hδ line was reported by Rivinius et al. (2006).

*HR 8539–π Aqr*. The binary nature of π Aqr was discovered by Bjorkman et al. (2002), who detected antiphased RV variations of the emission and absorption components of the Hα line. The primary CBe showed only weak signs of the disk presence at the time, with a traveling emission component originating from the vicinity of the secondary component in the outer disk and tracing its orbital motion. The authors also interpreted the observed long-term variations as an indication of variable mass exchange between the components. Zharikov et al. (2013) showed that the V/R variations are phase-locked with the binary orbit. Both studies conclude that the companion is unlikely to be an sdOB star, since its mass most likely exceeds ∼2 $M_\odot$. Due to its X-ray properties being similar to those of γ Cas, and the fact that the likely nondegenerate companion cannot be responsible for emitting X-rays, it is considered to be a member of the class of "γ Cas objects" (Nazé et al. 2017).

*HR 8762–o And*. The B component of this quadruple system is a double-lined spectroscopic binary that seems to be composed of two B-type stars of similar masses, one of which is the CBe (Hill et al. 1988; Zhuchkov et al. 2010).

### A.2. Wide Binaries

*HR 472–α Eri*. The brightest and closest CBe is an eccentric binary with a component separation ranging from 0″.05 to 0″.3 (Dalla Vedova et al. 2017; Kervella & Domiciano de Souza 2007; Kervella et al. 2008). It has an orbital period of ∼7 yr with the last periastron having occurred in 2015 (P. Kervella et al. 2019, in preparation). The primary entered a Be phase in 2013, indicating that the presence of the companion has no connection with the disk growth (Dalla Vedova et al. 2017).

*HR 2356, 2357, 2358–β Mon ABC*. The triple system β Mon consists of three CBes of similar brightness and spectral type. However, the B component has a much lower projected rotational velocity and also very little line emission, indicating only a weak disk presence (Cowley & Cowley 1966; Cowley & Gugula 1973). The components are separated by approximately 7″.1, 9″.7, and 2″.6 for the AB, AC, and BC pairs, respectively (Mason et al. 2001).

*HR 4621–δ Cen*. A companion contributing ∼7% in the K band (likely B- or A-type star) separated by 68.7 ± 0.5 mas was found by VLTI/AMBER spectro-interferometry, with the lower limit on the orbital period being ∼4.6 yr (Meilland et al. 2008).

## Appendix B
## Archival and Previously Published Data

### B.1. Archival IR Space Photometry

#### B.1.1. IRAS

*IRAS* was a space telescope that conducted an all-sky survey in four IR wavelength bands centered at 12, 25, 60, and 100 μm. The mission ceased operations in 1983 November after having surveyed more than 96% of the sky. The limiting flux density was about 0.5 Jy for the 12, 25, and 60 μm bands and 1.5 Jy for the 100 μm band. With a primary mirror size of 60 cm, the diameters of 80% of the encircled energy on the sky were 25″ at 12 and 25 μm, 60″ at 60 μm, and 100″ at 100 μm. The positional accuracy was reported to be usually better than 20″ (Beichman et al. 1988).

The measurements for the present study were taken from the Vizier *IRAS* point-source catalog II/125 (Helou & Walker 1988). We carefully selected only good-quality measurements and took special care to discard those potentially affected by confusion with other IR sources or contaminated by overlapping extended regions. The full selection criteria also included a signal-to-noise ratio (S/N) of at least 5 and a strong confidence of the point-source structure of the detected CBes.

#### B.1.2. AKARI

*AKARI* conducted an all-sky survey at IR wavelengths between 2006 May and 2007 August. It had two instruments on board: the infrared camera (IRC) covering 2–26 $\mu$m and the Far-Infrared Surveyor (FIS) covering 50–200 $\mu$m. The all-sky survey consisted of measurements centered at 9 and 18 $\mu$m (IRC) and 65, 90, 140, and 160 $\mu$m (FIS). *AKARI* was more advanced than *IRAS* in practically all aspects of the measurements, for instance, a sensitivity of ∼0.05 Jy, flux accuracy of 2%–3%, target position accuracy of 0″.8, and beam size of 9″ for the IRC. For more details about the *AKARI* all-sky survey, we refer the reader to Ishihara et al. (2010).

The data were extracted from the Vizier catalogs II/297 (Ishihara et al. 2010) and II/298, which correspond to the IRC and FIS instruments, respectively. Special care was taken to select only reliable data that were not affected by issues such as confusion or contamination by extended regions. This resulted in only a handful stars having usable data from the FIS, mostly in the 90 $\mu$m band.

### B.1.3. WISE

*WISE* was an IR all-sky space mission that was launched in 2009 December and ended its primary mission in 2011 February. It surveyed the sky in four broad bands with effective wavelengths of 3.4, 4.6, 11.6, and 22.1 $\mu$m. The photometric sensitivity was 0.12, 0.16, 0.85, and 4.0 mJy in the four bands, respectively, with a photometry accuracy of 2.4%–3.7% across the four bands and an astrometric accuracy of <0″.5. More information can be found in the official explanatory supplement (Cutri et al. 2012).

The *WISE* detections of the program stars were extracted from the Vizier catalog AllWISE data release II/328 (Cutri et al. 2013). Similar to the previous two missions, we took extensive care to only select high-quality data not affected by the usual problems, such as confusion with other IR targets, contamination by extended regions, low S/N, and fluxes reported as likely variable.

## B.2. Radio Submillimeter/Millimeter Flux Density Measurements Taken from Literature

### B.2.1. IRAM 30 m

The 30 m telescope of the Institute for Radio Astronomy in the Millimetre range (IRAM), one of the most sensitive single-dish telescopes in the millimeter domain, is located in the Spanish Sierra Nevada (Baars et al. 1987). The data used in this study were obtained with a bolometer detector system centered at 250 GHz ($\lambda = 1.2$ mm) with a beamwidth of 11″.25 ± 0″.15 (Kreysa 1990) and published by Altenhoff et al. (1994) and Wendker et al. (2000).

### B.2.2. JCMT

The James Clerk Maxwell Telescope (JCMT) is a 15 m submillimeter/millimeter telescope located on Maunakea in Hawaii. The observations used in this study were taken using the now-retired UKT14 bolometer at central wavelengths of 0.8 and 1.1 mm with half-power beamwidths of 16″ and 18″.5 respectively (Duncan et al. 1990). More details on the observations are given by Waters et al. (1989, 1991), who published the measurements used in this study.

## B.3. Radio Centimeter Flux Density Measurements Taken from Literature

### B.3.1. VLA

The VLA is situated in New Mexico and consists of 27 antennas with 25 m diameters distributed in a Y-shaped configuration. The VLA was in operation from 1970 to 2010, when it was vastly upgraded and renamed the Karl G. Jansky VLA. We used previously published VLA flux density measurements, as presented in the works of Taylor et al. (1987, 1990), Apparao et al. (1990), Dougherty et al. (1991), and Dougherty & Taylor (1992).

### B.3.2. ATCA

The Australia Telescope Compact Array (ATCA) is located in eastern Australia and consists of six 22 m antennas. The data used for this study were published by Clark et al. (1998) and consist of measurements of 13 CBes at central wavelengths of 3.5 and 6.3 cm. We note that the detection of $\mu$ Cen was tentative and has not yet been confirmed.

## Appendix C
## SED Analysis—Detailed Discussion of Individual Targets

### C.1. Stable and Weakly Variable CBes with SED Turndown

#### C.1.1. Known Close Binaries

*HR 264–$\gamma$ Cas*. This spectroscopic binary star was analyzed in detail by Klement et al. (2017b). One upper limit measured by the VLA suggests a possible variability of the radio fluxes. The spectral slopes in the IR and centimeter are remarkably similar but with the radio part shifted to lower fluxes. We interpret this as a sign of the circumbinary nature of the CBe disk.

*HR 1910–$\zeta$ Tau*. The IR data from all three missions show good agreement for this spectroscopic binary CBe. The H$\alpha$ profile has shown complex variations, including periodic V/R changes and additional emission peaks appearing (see Štefl et al. 2009, for a summary). The one-armed density wave propagating through the disk was studied in detail by Carciofi et al. (2009) and Escolano et al. (2015). The emission-to-continuum ratio (E/C) of the stronger peak was measured to be 2.25 in 1972 (Gray & Marlborough 1974) and 1.8 in 1980 (Andrillat & Fehrenbach 1982). Throughout the 1980s and early 1990s, the E/C initially decreased from 3 (1982) to 2 (1987) before rising to 4 (1993). BeSS spectra show that during *AKARI* observations in 2006/2007, the emission was rather variable, with an E/C between 2 and 4, while throughout 2010 (*WISE* epoch), it stayed at around 2.5. Since then, the emission has further decreased to 1.5–2.0, with strong V/R variations dying down. Rich data sets in both the millimeter and centimeter regions show evidence for variability in the outer disk as well. Older data from the IRAM 30 m (1991–1992) show a detection, while new LABOCA and CARMA data (both 2013) around 1 mm show only upper limits (3$\sigma$ level) that are inconsistent with the IRAM 30 m detection. A CARMA measurement at 3 mm taken 5 days later, however, shows a detection, which seems to rule out a sharp SED turndown. Most interestingly, the new 2010 JVLA measurements taken 6 days apart show a flux at 5 cm, which is similar to the upper limit derived at 3.54 cm, suggesting a flat or even positive spectral slope in this region. This behavior indicates a complex structure of the outer disk, likely caused by both the density wave and the complicated nature of disk truncation by the binary companion. In any case, all radio measurements lie clearly below the extrapolation of the IR spectral slope.

*HR 2142*. This confirmed Be+sdOB binary system shows IR fluxes that are consistent across the three missions. The H$\alpha$ line profile shows variations in its shape, with a double-peaked appearance often changing to a complex triple-peaked one. The overall emission, however, has remained very stable since at least 1982, with E/C ≈ 3.5–4 (Andrillat & Fehrenbach 1982; Hanuschik et al. 1996). An SED turndown is noticeable in the centimeter region, where the 3$\sigma$ upper limit from the VLA lies about 1$\sigma$ below the extrapolation of the combined IR slope.

*HR 2845–β CMi*. This remarkably stable CBe is now a confirmed spectroscopic binary (Dulaney et al. 2017). It has been studied in great detail using radiative transfer modeling by Klement et al. (2015, 2017b), who first suggested the presence of an unseen binary companion due to the observed SED turndown, which in the VDD model could not be explained otherwise. Here we present new radio detections from the JVLA at wavelengths of 1.36, 3, 5, and 10 cm, which agree very well with the lower-S/N detections from the past, confirming the stability of the disk including its outer parts over a timescale of more than 30 yr. As the disk of $\beta$ CMi is very tenuous, it is probably more appropriate to fit the slope of both the IR and millimeter data in order to search for SED turndown, as was done using physical models by Klement et al. (2015, 2017b). Nevertheless, the change of spectral slope between the IR and radio is very apparent thanks to the wealth of radio data, which include measurements from LABOCA, JCMT, IRAM 30 m, VLA, and JVLA. As derived from a physical model by Klement et al. (2017b), the radio slope is indicative of a gradual truncation, or perhaps a circumbinary disk rather than a sharp truncation. The circumbinary scenario will be explored in further detail in a future publication.

*HR 4787–κ Dra*. This confirmed binary shows excellent agreement between IR measurements from all three missions. In the past, the double-peaked H$\alpha$ emission showed an E/C varying between ∼2 and ∼3, as seen in the few spectra measured in 1972, 1980, and 1999 (Gray & Marlborough 1974; Andrillat & Fehrenbach 1982; Banerjee et al. 2000). The available BeSS spectra show a higher E/C of ∼3.5 and, often, a more complex line profile shape between 2006 and 2011. Since then, the E/C slowly decayed to 2.5 in 2015 and continued to 1.5 in 2019. The SED turndown is detected at millimeter wavelengths and more strongly at centimeter wavelengths, as revealed by the upper limits measured by the IRAM 30 m and the VLA. Given the lasting presence of the disk, albeit somewhat variable, the SED turndown is likely due to truncation by the binary companion on a 61.6 day orbit (Section 3.1.1).

### C.1.2. Unconfirmed Close Binaries

*HR 1087–ψ Per*. The second-brightest CBe at radio wavelengths, $\psi$ Per is the first to be detected at centimeter wavelengths (Taylor et al. 1987). The SED turndown is very apparent, and old VLA data suggest a possible variability of the outer disk. For detailed discussion and physical modeling, we refer the reader to Klement et al. (2017b).

*HR 1165–η Tau*. A clear turndown is detected for this star with plenty of radio data, which were discussed in detail in the context of radiative transfer modeling by Klement et al. (2017b). Since no new data are presented here, we refer the reader to that study for more detailed discussion.

*HR 1273–48 Per*. All three IR missions for this star show a remarkable agreement, indicating a very stable disk present since 1983. An H$\alpha$ profile measured in 1980 shows an E/C of ∼5 (Andrillat & Fehrenbach 1982), and in 1998, there was an E/C of ∼7 (Banerjee et al. 2000). The available BeSS spectra show an E/C of 6–7 from 2007 to 2019. A clear turndown in the SED is revealed by the strong upper limit measured by the VLA.

*HR 1622–11 Cam*. The *WISE* and *AKARI* data are in close agreement, with the exception of an outlying *AKARI* point at 90 $\mu$m. The one *IRAS* data point lies slightly above the *WISE* and *AKARI* data, indicating possible small-scale variability. In the early 1970s, the E/C of the H$\alpha$ line was reported to be 6.43 (Gray & Marlborough 1974). In 1980, it was close to 6.2 (Andrillat & Fehrenbach 1982), while in 1983, it apparently rose to ∼10 (Ballereau et al. 1987), which helps explain the higher flux level measured in 1983 by *IRAS*. By 1999, the E/C was back to ∼6 (Banerjee et al. 2000). The available spectra in the BeSS database show a reasonably stable H$\alpha$ profile with a mild variability of the E/C between 5 and 7 in 2007–2019. The available upper limit from the VLA lies well below the extrapolation of the IR data, even when considering the mild variability.

*HR 1660–105 Tau*. The IR data from the three missions are in good agreement, with the exception of a lower-quality *AKARI* flux at 90 $\mu$m. An H$\alpha$ profile from 1981 shows an E/C of ∼5 (Andrillat & Fehrenbach 1982), while between 1988 and 1999, it was varying between 6 and 7 (Hanuschik et al. 1996; Banerjee et al. 2000). The BeSS spectra show a slightly increased emission with an E/C rising from ∼7.5 in 2007 to ∼9 in 2010. A clear SED turndown is observed thanks to the upper limit measured by the VLA.

*HR 1956–α Col*. The IR missions are in close agreement and show a spectral slope consistent with the LABOCA detection at 0.87 mm. The double-peaked H$\alpha$ line profile has shown an E/C close to 2.5 since at least 1982 (Hanuschik et al. 1996; Banerjee et al. 2000). A clear turndown in the SED is revealed by strong upper limits from both the VLA and ATCA. This star was recently modeled using the procedure described in Chapter 6 of Mota (2019), and the radio data indicate a truncation radius of ∼20 stellar radii (A. C. Rubio 2019, private communication).

*HR 2356 and 2358–β Mon A and C*. The interpretation of the SED of this triple CBe is complicated by the small separation of the components (<10″). The IR flux measurements from *IRAS* and *AKARI* therefore contain contributions from all three components, and it is not possible to disentangle them. Historically, the H$\alpha$ emission was very weak for the B component, indicating that only a very weak disk was present. Assuming that this is the case to this day, the radio upper limits measured for the B component do not indicate an SED turndown but rather a nonexistent or weak disk. If this is indeed the case, the IR data then consist of contributions only from the A and C components separated by 9″.7. The A component has shown slightly variable H$\alpha$ emission with an E/C close to 4, as shown by available spectra since 1980 (Andrillat & Fehrenbach 1982; Hanuschik et al. 1996; Banerjee et al. 2000), while the C component also showed an E/C close to 4, as seen in spectra from 1980 (Andrillat & Fehrenbach 1982) and 1998 (Banerjee et al. 2000). We can therefore simply assume an equal contribution to the IR fluxes from the A and C components and little to no contribution from the B component. The submillimeter/millimeter data correspond to the A component only, assuming a good pointing accuracy and negligible contribution from the B component separated by 7″.1. In addition to the historic radio centimeter data for the A component, we present new JVLA measurements (Table 6), which provided multiple detections for the A and C components, including a detection at 10 cm for the A component. When splitting the IR flux values evenly between the A and C components, we still detect a clear SED turndown for both. The SED plots for the three components, which are included in Figure Sets 2 and 3 (available in the online journal), contain IR measurements split between the A and C components in the way described above, with the B component being left with no IR measurements. Previously, $\beta$ Mon A was studied by Klement et al. (2017b), who found only weak evidence for the SED turndown due to the inclusion of WISE data in the IR SED, which indicated a steeper IR slope. However, we found that the WISE data are actually flagged as contaminated by scattered light from a nearby bright source, and we therefore discarded them from the present analysis.

*HR 4140–p Car*. Detections in only two WISE bands and one AKARI band are available for this star. The H$\alpha$ line profile has been V/R variable and occasionally showed a complex triple-peaked shape with an E/C close to 4 between 1985 and 1993 (Hanuschik et al. 1996) and between 2012 and 2017 (BeSS and BeSOS). We were not able to assess the variability of the disk at the times of the IR observations, as no spectra are available in the literature. The radio data set consists of one 2017 measurement from LABOCA. The LABOCA map shows complex, extended structures that do not originate from the CBe. We were able to extract an upper limit for p Car, which indicates a turndown in the SED between IR and millimeter wavelengths.

*HR 4621–δ Cen*. The object $\delta$ Cen is the brightest CBe at radio wavelengths, both millimeter (LABOCA) and centimeter (ATCA). Even so, a turndown with respect to the IR fluxes is apparent already at millimeter and much more strongly apparent at centimeter wavelengths. The H$\alpha$ line profile has shown a complex shape with an E/C of 7–9 between 1985 and 1993 (Hanuschik et al. 1996) and 7 in 1999 (Banerjee et al. 2000). A total of six spectra available at BeSS and BeSOS show E/C levels between 7 and 9.

*HR 5941–48 Lib*. A dense disk has been present for many decades, with the H$\alpha$ line exhibiting strong V/R variations indicative of a disk density wave with a period of more than $\sim$10 yr (see Silaj et al. 2016, and references therein). Since the star is seen close to edge-on, this may affect the IR measurements, as the denser part of the inner disk may be periodically shielded by the stellar photosphere. However, the IR data show a good agreement and confirm the overall stability of the average properties of the disk. Even though there is only one data point available from each IR mission, the SED turndown seems to be apparent already at millimeter wavelengths, where the LABOCA and IRAM 30 m detections lie below the extrapolation of the IR slope. The turndown is further confirmed by the $3\sigma$ upper limits measured by the VLA, as well as the new upper limit measured by the JVLA.

*HR 6118–χ Oph*. This CBe shows reasonable stability of the disk, as shown by overall consistent IR fluxes. It has shown very strong emission in H$\alpha$, with an E/C reported to be close to 20 in 1972–1973 (Gray & Marlborough 1974), but it apparently decreased to $\sim$9 by 1980 (Andrillat & Fehrenbach 1982). Between 1985 and 1993, the E/C was between 9 and 11 (Hanuschik et al. 1996), and the line profile showed a complex triple-peaked shape. In 1998, the E/C was measured to be close to 11 (Banerjee et al. 2000), and according to BeSS spectra, it has fluctuated between 10 and 17.5 since. From the limited number of measurements, it seems that at the time of the AKARI and WISE observations, it was more likely at the higher end of this interval. The IR fluxes themselves indicate that the disk was denser at the time of the WISE observations than at the time of the AKARI observations, while IRAS reflects a more tenuous state during the 1980s. The SED turndown is very clearly revealed when comparing old IRAS data with the old upper limits from the VLA and ATCA, as well as when comparing the newer IR data sets with the LABOCA detection from 2017. The object $\chi$ Oph has been proposed to be a close binary, but this was never confirmed (Section 3.1.1).

*HR 6510–α Ara*. The IR data from all three missions are in good agreement, with the exception of an outlying AKARI point at 60 $\mu$m. This measurement is quite suspicious, as it differs remarkably from the other AKARI detections at 18 and 90 $\mu$m. Between 1982 and 1993, the H$\alpha$ emission remained reasonably stable with an E/C of 3.5–4.5 (Hanuschik et al. 1996), and a spectrum from 1999 shows an E/C of $\sim$3.3 (Banerjee et al. 2000). More recent spectra from the BeSS and BeSOS online databases show an E/C rising from $\sim$4 to $\sim$5.5 between 2008 and 2018. A clear turndown is revealed already in the millimeter region thanks to LABOCA detection and is further confirmed in the centimeter region by the upper limits measured by ATCA. The object $\alpha$ Ara was recently modeled by Mota (2019), who found that the millimeter data are consistent with a truncation radius of $16^{+4}_{-3}$ stellar radii.

*HR 8402–o Aqr*. The IR measurements are in good agreement, and, from the available measurements, the double-peaked H$\alpha$ profile has remained reasonably stable as well. A measurement from 1976 shows an E/C of 2.92 (Fontaine et al. 1982), and BeSS spectra show an E/C rising from $\sim$2.8 (in 1995) to 3–3.5 (2001–2018). There is somewhat weak evidence for an SED turndown in the millimeter region, as shown by one of the two upper limit measurements from the IRAM 30 m. Here we present one new upper limit and three detections measured by the JVLA in 2010 that lie slightly below the old upper limit measured by the VLA. The centimeter data clearly indicate a strong turndown in the spectral slope.

*HR 8731–EW Lac*. This star with a wealth of radio data was studied in detail by Klement et al. (2017b). The radio data in the centimeter region indicate variability of the outer disk, which may be connected to density waves propagating through the disk, as seen in the V/R variations of emission line profiles. However, a clear turndown is still revealed.

### C.2. CBes with Strong Disk Variations

#### C.2.1. Variable CBes with Evidence for SED Turndown

*HR 496–φ Per*. This confirmed Be+sdOB binary exhibits a variable H$\alpha$ line profile with an E/C of 4.58 in 1972 (Gray & Marlborough 1974), 5–7 in 1980 (Andrillat & Fehrenbach 1982), and 3 in 2000 (Banerjee et al. 2000). BeSS spectra from 1993 to 2019 show a frequent appearance of additional peaks in the emission profile and an E/C oscillating between 3 and 5.

During *AKARI* and *WISE* measurements, it was close to ∼4 and ∼3, respectively. Even given the history of moderate disk variability and correspondingly varying IR fluxes, the radio detection from the IRAM 30 m and the upper limit from the VLA (taken in 1992 and 1988, respectively) indicate a clear SED turndown, as is expected from a close binary companion truncating the disk.

*HR 1423–228 Eri*. The IR measurements suggest a stronger disk present during the early 1980s than in the late 2010s. Indeed, the Hα emission had an E/C of 2.8–3.2 between 1982 and 1989, with the emission collapsing to 1.3 by 1993 (Hanuschik et al. 1996). Therefore, we are able to compare *IRAS* data with the VLA upper limit taken in 1987 while assuming a stable disk during that period, and a clear SED turndown is revealed. BeSS spectra show that during 2006–2007, the E/C increased from 2.3 to 2.6, and in 2010, it was 2.6 as well. The slightly lower E/C compared to the spectra from the 1980s explains the lower flux densities measured by the later IR missions compared to *IRAS*. Since 2010, the E/C of Hα rose to ∼4.5 at the end of 2018 while showing complex features in the line profile.

*HR 2538–κ CMa*. The IR data indicate a weaker disk present during *AKARI* observations as compared to the other two missions. The Hα line profile shows a complex shape with an E/C of 4–5.5 during 1982–1993 (Hanuschik et al. 1996), while by 1999, it was close to ∼3 (Banerjee et al. 2000). BeSS spectra show an E/C of 4–4.5 in 2006–2007, while by 2010, it rose back to ∼5, thus confirming what was inferred from the IR measurements. The disk remained strong at least until early 2019. Given the observational history, we are able to confidently detect an SED turndown between *IRAS* and the VLA upper limit (1988), as well as between *AKARI*/*WISE* and the LABOCA upper limit (2018).

*HR 2911–OW Pup*. Two available Hα spectra from 1985 and 1992 show a double-peaked profile with E/Cs of ∼5.5 and ∼5, respectively. BeSS and BeSOS spectra from 2012 to 2013 show an E/C of 3.5, while by 2019, the emission had declined to ∼2. Assuming overall stability of the disk during the 1980s, the VLA upper limit (1987) likely indicates an SED turndown.

*HR 6712–66 Oph*. Between 1972 and the turn of 1976/77, the E/C of Hα was reported to be between 3.2 and 3.7 (Gray & Marlborough 1974; Slettebak & Reynolds 1978; Fontaine et al. 1982). After that, the emission was gradually increasing until reaching an E/C of 10 in 1993 (Andrillat & Fehrenbach 1982; Apparao et al. 1990). The disk then continued to monotonically decay until it completely disappeared in 2011 (Miroshnichenko et al. 2012). It has remained in a diskless state since. The great difference between the IR fluxes of 66 Oph measured by *IRAS* and the later missions is consistent with the history of disk presence: the *IRAS* fluxes reflect a well-developed disk, while *AKARI* and *WISE* show more advanced stages of the disk decay. The IRAM 30 m detection and VLA upper limit were measured close to the maximum strength of the disk in the late 1980s and early 1990s and indicate a clear turndown with respect to the *IRAS* fluxes. ATCA measurements from the late 1990s show upper limits consistent with the SED turndown, but they may already reflect disk dissipation. Interestingly, the 2010 data from JVLA show three clear detections at centimeter wavelengths at similar values as the 3σ upper limits from the VLA and ATCA. The detected fluxes can be interpreted as originating from the remnants of the outer disk, which at that time was not yet fully dissipated. In that case, however, the ATCA measurements should have shown higher fluxes, reflecting the disk dissipation in early phases. The disparity between the ATCA and JVLA fluxes may indicate variability in the outer disk, where the presence of an unseen companion possibly complicates the simple picture of CBe disk dissipation.

*HR 8773–β Psc*. Available Hα spectra show an E/C of 7 in 1982, 4.5 in 1987, and 6.5 and 8.5 in 1989 and 1993, respectively (Hanuschik et al. 1996). BeSS spectra show an E/C of 5 in 1992–1993, rising to 6 in 1995. The following measurement from 2001 shows an E/C of 6.5. Assuming no strong variations between 1995 and 2001, *IRAS* and ATCA observations took place when the emission had a similar strength with an E/C of 6–7, and the SED shows a strong turndown between the *IRAS* detections and ATCA upper limits. The *AKARI* fluxes indicate a stronger disk than the *IRAS* fluxes, but the available spectra show an ambiguous picture with Hα from late 2004 showing an E/C of ∼7.5, followed by an E/C of ∼4 in the second half of 2007. In late 2010, the emission was back at around 7.3–8.0, and it remained between 6 and 8 at least up to early 2019. About a month after our LABOCA measurement, the E/C was ∼8. The comparison of the spectral slope derived from the *WISE* IR data (2010) with the LABOCA detection (2017) does not indicate a turndown in the SED.

*C.2.2. Variable CBes with Inconclusive Data*

*HR 193–o Cas*. According to the IR data, a weaker disk was present during *IRAS* observations as opposed to *WISE* and *AKARI* data, indicating a stronger and stable disk during the late 2010s. Indeed, an Hα spectrum from 1980 shows an E/C of ∼1.9 (Andrillat & Fehrenbach 1982), while BeSS spectra from 2006–2010 show an E/C of 6–7. The closest available Hα measurement to the VLA upper limit from 1988 was taken in 1993 and shows an E/C of ∼5. An SED turndown is not detected.

*HR 2170*. Based on the IR data, the disk was well developed during the *IRAS* measurements, less so during the *AKARI* measurements, and dissipating during the *WISE* measurements. Assuming that the disk was stable between the *IRAS* epoch (1983) and the VLA upper limit measurement (1987), we are detecting a strong turndown in the SED. Unfortunately, we were unable to find contemporary spectra in the literature, so this cannot be confirmed. BeSS spectra from 2012 to 2018 show an E/C of 3.0–3.5.

*HR 5193–μ Cen*. The disk surrounding μ Cen completely disappeared between 1973 and 1976, with Hα emission collapsing from an E/C = 5 to zero. Throughout the 1980s, μ Cen showed flickering activity and occasional outbursts, with four larger outbursts taking place during 1987 alone (see Figure 1 of Hanuschik et al. 1993). A new disk started building up in 1992 with an E/C reaching 1.5–2 in 1993 (Figure 55 of Hanuschik et al. 1996). The available BeSS and BeSOS spectra from 2014 to 2017 show stronger emission with an E/C of 3.5–4. The IR measurements suggest the presence of a disk even at the time of the *IRAS* observations in 1983 and a much denser disk during the *AKARI* observations. On the other hand, *WISE* measurements suggest a tenuous disk corresponding to a likely dissipation phase during 2010. We note that the ATCA detection at 3.5 cm is unconfirmed (see Section 3 of Clark et al. 1998). Whether real or not, an SED turndown is not observed and, in any case, would be difficult to confirm given the particularly complicated history of disk presence around this star.

*HR 5440–η Cen*. Another strongly variable disk showing Hα in pure absorption in 1982 and an E/C rising to ∼1.7 by 1987, while between 1987 and 1993, it went steadily down to ∼1.2 (Hanuschik et al. 1996). In 1999, it was close to 1.4 (Banerjee et al. 2000). In 2006, the emission was at ∼2.1 (T. Rivinius 2019, private communication), which explains the higher IR fluxes detected by *AKARI*. Unfortunately, the few spectra available in BeSS and BeSOS cover only the period between 2014 and 2017, again showing a reduced E/C of 1.2 to 1.5. *WISE* data seem to indicate a dissipating disk intermediate between the *AKARI* and *IRAS* level. The radio data set consists of two detections by LABOCA from 2013 and 2015, which both lie below the extrapolation of *AKARI* and *WISE* data. However, given the disk variability, we cannot confidently interpret this as a sign of SED turndown.

*HR 6175–ζ Oph*. This star was observed in the IR during a diskless phase, which is confirmed by very little or no line emission present in BeSS spectra from 1995 to 2019. We found only one archival spectrum from 1976 in the literature, and it also shows Hα in absorption. Following an upper limit measurement by the IRAM 30 m from 1989 (Altenhoff et al. 1994), ζ Oph was detected by the same telescope in 1992 with a flux exceeding the previous upper limit by $2\sigma$ (Wendker et al. 2000). Given the lack of contemporaneous spectra, we are unable to confirm whether a disk was actually present during the early 1990s. In any case, an SED turndown is not detected.

*HR 6304*. Available Hα spectra show a complex multi-peaked profile in 1989 with an E/C of ∼4.5 and symmetric double-peaked profiles with an E/C of ∼3.5 in 1992 and 1993 (Hanuschik et al. 1996). We found no other spectra in the literature. Three spectra available in the BeSS database show a single-peaked Hα profile with an E/C rising from ∼4.5 to ∼6.5 between 2012 and 2014. The IR data show a stronger disk during the more recent measurements. Assuming that the disk was stable between 1997 (ATCA) and 2006–2007 (*AKARI*), we see evidence of a turndown based on one of the upper limits from ATCA.

*HR 6451–ι Ara*. This candidate Be+sdOB system (Wang et al. 2018) shows a stronger disk presence during *IRAS* than during *AKARI* and *WISE*, which show a good agreement with each other. The radio data set consists of only one upper limit from LABOCA measured in 2017, which indicates SED turndown only with respect to the *IRAS* detections from 1983. Available Hα spectra show a double-peaked profile declining from an E/C of ∼4.5 to ∼3.5 from 1985 to 1993, with the exception of a 1987 spectrum showing a complex triple-peaked profile (Hanuschik et al. 1996). BeSS spectra show a varying profile shape and an E/C between 2.5 and 3.5 in 2010–2018.

*HR 7708–28 Cyg*. This candidate Be+sdOB binary shows an outlying *IRAS* point at 12 μm, indicating that a much stronger disk was present in the early 1980s compared to the late 2000s. Unfortunately, no spectra are available to verify this. However, a measurement from 1976 shows an Hα E/C of 2.18 (Fontaine et al. 1982), and BeSS spectra taken between 1995 and 2005 show an E/C close to 2.5, after which the emission decreased to ∼1.3 by 2007 and further to ∼1.1 by 2011. Soon after, the emission started slowly rising back to the current level of ∼2. With only one *IRAS* detection available, we are unable to estimate the spectral slope in the IR before the disk decay during the *AKARI* and *WISE* observing epochs.

Therefore, the possible presence of an SED turndown cannot be confirmed.

*HR 7807*. This candidate Be+sdOB binary has only *AKARI* and *WISE* IR data available. Using only the mid-IR *AKARI* data for the IR spectral slope fit, an SED turndown with respect to the old IRAM 30 m data is revealed. However, we were not able to find any archival Hα measurements taken close to when the radio data were obtained (1992); therefore, we cannot verify if a disk was present or how strong it was. BeSS spectra show an E/C of 1.7–2.0 between 2001 and 2015; after that, the emission gradually decreased to 1.2 in 2018.

*HR 8053–60 Cyg*. According to the available BeSS spectra, the confirmed Be+sdOB binary 60 Cyg had a developed disk in 1999 (Hα E/C = 1.5) that completely disappeared by 2001. A new disk started building up between 2004 and late 2007 and remains to this day, with an E/C fluctuating between ∼1.7 and ∼2.2. Based on the *AKARI* and *WISE* fluxes, a weaker disk was present during 2006–2007 than in 2010, when an E/C of Hα was around 1.75. Given the disk variations and limited radio data set, we are unable to detect an SED turndown.

*HR 8260–ε Cap*. This binary CBe has an outlying radio detection by the IRAM 30 m in the millimeter region, suggesting a stronger disk presence during the early 1990s compared to both the preceding epoch (*IRAS* detection) and afterward (*WISE*, *AKARI*, JVLA upper limits). Available Hα spectra show an E/C rising from 1.3 in 1982 to 1.6 by 1989, after which the disk declined to 1.2 by 1993 (Hanuschik et al. 1996). BeSS spectra show an E/C rising from 1.2 in 2001 to 1.7 by 2008 and staying there until 2011 before declining to 1.3 by 2017. Given this history, the outlying millimeter detection is quite peculiar and indicates that the disk continued getting stronger from 1989 to 1992 before declining in 1993. Only upper limits could be measured from our new JVLA data (Table 6). Weak evidence for an SED turndown is provided by the upper limit at 5 cm when considering the IR slope from the *AKARI* detections only.

*HR 8375*. The IR continuum fluxes show a strong disk presence during *IRAS* measurements but no evidence for a disk during the *AKARI* and *WISE* observations. A clear turndown between the *IRAS* points (1983) and the VLA upper limits (1988) is present. The Hα E/C was close to 5 in 1976 (Fontaine et al. 1982), but unfortunately, we lack any further measurements up until 2001, when a BeSS spectrum shows an E/C of only ∼1.2. A weak emission component on top of the photospheric absorption profile (E/C = 0) was still present in 2003, but by 2006, the emission disappeared completely.

### C.3. Special Cases

*HR 472–α Eri*. This star has shown recurrent phases of disk buildup and dissipation, which was suggested to be cyclic with a period of 14–15 yr (Vinicius et al. 2006). More recent history suggests no clear periodicity on the basis of the latest outburst starting in 2013 after a diskless state lasting ∼7 yr (Faes 2015; Faes et al. 2015). The history of disk presence is well sampled by the IR observations: *IRAS* data were taken during a diskless state, *AKARI* data during the late stages of disk dissipation, and *WISE* data again during a diskless state. A few months prior to the latest periastron passage in 2015 November, the disk started dissipating, but it recovered and is present to this day. The LABOCA observations were executed over 5 consecutive nights during the later stages of the periastron passage and about 5 months after. The fluxes measured during the periastron are dominated by disk contribution and show day-to-day variations with a noticeable dip on the second night, after which the flux got back to its original level on the third night. The fluxes were more stable over the last 2 of the 5 consecutive nights, suggesting an initial perturbation of the outer disk stabilizing over time. The post-periastron flux density is consistent with the measurements taken during periastron. Studies dedicated to the binary orbit and periastron passage are in preparation.

*HR 2749–ω CMa.* This remarkable CBe has gone through four full cycles of disk dissipation and buildup during the last four decades. The complicated history and its implications for the nature of viscosity operating in astrophysical disks are best summarized by Ghoreyshi et al. (2018). The *V*-band light curve analyzed in that work corresponds to emission coming from a region very close to the star only and therefore represents the best probe into a varying mass-loss rate at the base of the disk. The IR continuum and Hα emission, on the other hand, both originate from larger disk areas and represent more averaged properties of the disk with a delayed response to the varying conditions in the inner disk. Indeed, Hα emission shows an increase from an E/C of ~3.5 in 1982 to ~6 in 1989 and back to 3.5 by 1993 (Hanuschik et al. 1996), while in the *V*-band light curve, 1989 actually corresponds to the middle of a quiescent phase. In 1999, the E/C of Hα was close to 7 (Banerjee et al. 2000), while BeSS spectra show an E/C of 6–7 in 2006–2007 decreasing to 4 in late 2008. More recent spectra continue to oscillate between an E/C of 4 and 6.5. The IR continuum fluxes show similar behavior to Hα: during the *IRAS* and *WISE* observing epochs, the disk was less developed, while in 2006–2007, the *AKARI* data point at 18 μm shows enhanced emission, even though the *V* band shows a quiescent phase at that time. We took two LABOCA measurements in 2015 and 2017, corresponding to the end of a quiescent phase and the beginning of a new growth phase, respectively, and were able to detect a decrease in the submillimeter flux from ~23 mJy to a 3σ upper limit of 15 mJy. This will be detailed along with the evolution of other observables in a future dedicated paper. Given the complex nature of the disk of ω CMa, we are unable to reach a firm conclusion regarding a possible SED turndown.

*HR 5953–δ Sco.* Weak Hα line emission was first detected by Cote & van Kerkwijk (1993) at a time close to the 1990 periastron passage. The emission remained weak until 2000, when the primary B-type star component of δ Sco started steadily developing a circumstellar disk, as documented by rising emission in Hα, as well as overall brightening in the visible and IR regions (Carciofi et al. 2006; Miroshnichenko et al. 2013). Since then, the disk has remained stable overall with an E/C in Hα of 3–5. This history is reflected in the IR data: the *IRAS* measurements were taken during the diskless state, while the *AKARI* and *WISE* measurements correspond to epochs when the disk was already well developed. The radio measurements by LABOCA and CARMA were taken before and after the latest periastron passage in 2011. Both clearly show a decrease in the observed radio fluxes following the periastron, which indicates disruption of the outer disk by the close approach of the companion. The effects of the periastron passage were also observed in the profile of Hα, indicating that not only the outer disk was perturbed (Miroshnichenko et al. 2013). Multitechnique studies detailing the latest periastron passage are in preparation.

## ORCID iDs

Robert Klement https://orcid.org/0000-0002-4313-0169
A. C. Carciofi https://orcid.org/0000-0002-9369-574X
R. Ignace https://orcid.org/0000-0002-7204-5502
D. Gies https://orcid.org/0000-0001-8537-3583
N. D. Richardson https://orcid.org/0000-0002-2806-9339
C. de Breuck https://orcid.org/0000-0002-6637-3315
P. Kervella https://orcid.org/0000-0003-0626-1749